%% file: main.tex
\documentclass[10pt]{jfm}
\usepackage{amssymb,amsbsy, graphicx, booktabs, array, bbold, color,fancyhdr, empheq, natbib,subcaption, calligra, bm} 

\usepackage[a4paper,margin=1in]{geometry}
\usepackage[dvipsnames]{xcolor}
\usepackage[hidelinks]{hyperref}
\usepackage[nameinlink]{cleveref}

\usepackage{ulem}
\usepackage[colorinlistoftodos]{todonotes}

\input{SPTmacro}

\shorttitle{Elastohydrodynamics of 3D chemically active filaments}
\shortauthor{M. Butler, B. Walker, T. Montenegro-Johnson, P. Katsamba}

\title{Elastohydrodynamics  
of 3D chemically active filaments}
\author{
Matthew D. Butler\aff{1,2}\corresp{\email{matthew.butler@strath.ac.uk}},
Benjamin J. Walker\aff{1}, 
Thomas Montenegro-Johnson\aff{3}, 
Panayiota Katsamba\aff{4,5}\corresp{\email{panayiota.katsamba@cut.ac.cy}}}
\date{\today}

\affiliation{\aff{1}Department of Mathematics, University College London, London, WC1H 0AY, UK
\aff{2}Department of Mathematics \& Statistics, University of Strathclyde, Glasgow, G1 1XH, UK
\aff{3} Mathematics Institute, University of Warwick, Coventry, CV4 7EZ, UK
\aff{4}Computation-based Science and Technology Research Center (CaSToRC), The Cyprus Institute, Nicosia, 2121, Cyprus
\aff{5}Department of Chemical Engineering, Cyprus University of Technology, 30 Archbishop Kyprianou Str., Limassol 3036, Cyprus}

\begin{document}

\maketitle

\begin{abstract}
    Active deformable filaments exhibit a large range of qualitatively different three-dimensional dynamics, depending on their flexibility, the strength and nature of the active forcing, and the surrounding environment. 
    We investigate the dynamic behaviour of elastic, chemically propelled phoretic filaments, combining two existing models; a local version of
    slender phoretic theory, which determines the resulting slip flows for chemically propelled filaments with a given shape and chemical patterning, is paired with a computationally efficient method for capturing the elastohydrodynamics of a deformable filament in viscous flow to study the chemoelastohydrodynamics of filaments. 
    As the activity increases, or equivalently the filament stiffness decreases, these filaments undergo buckling instabilities that alter their behaviour from rigid rods. 
    We follow their behaviour well beyond the buckling threshold to find a rich array of dynamics. 
    Through two illustrative examples, we conduct initial-value simulations that show that, as the stiffness of the filament is decreased, the dynamic behaviour moves from rigid motion to planar buckling, through an out-of-plane transition, eventually reaching diffusive-like behaviours  for very deformable filaments. 
\end{abstract}


\section{Introduction}
\label{sec:Intro}

Active colloids have a prominent position as self-propelling agents that enable the study of multi-agent systems, called `active matter' \citep{ramaswamy2010mechanics}. These canonical systems of active matter serve as micro-engineered analogues of crowds, flocks of birds or bacterial swarms. 
A common propulsion mechanism is phoresis, where particles move in gradients of a surrounding field \citep{Anderson1989}. This occurs due to short-range interactions between the particle's surface and the surrounding field that generate an effective slip flow close to the particle's surface, such as solute concentration \citep{GolestanianLiverpoolAjdari2005}, temperature \citep{bickel2013flow} or electric field \citep{nourhani2015self}. If the particle also creates gradients in the surrounding propulsive field, then it can self-propel; this self-propulsion is called autophoresis \citep{Paxton2004}.

One class of autophoretic particles are chemically active (diffusiophoretic) colloids, which are microscale particles that swim by converting a solute `fuel' in their environment into propulsive slip flows \citep{GolestanianLiverpoolAjdari2007}. The canonical chemically active colloid is the `Janus' particle, which is half active and half inert; typically  this takes the form of a rigid sphere, rod, or disk, partially coated in a catalyst, such as platinum, in a surrounding solute such as hydrogen peroxide solution \citep{Howse2007,ebbens2011direct}. Differential reaction rates at the particle's surface create surface gradients in the solute concentration; the resulting short-range solute-surface interactions generate pressure gradients in a thin layer that drives slip flows close to the surface that propel the swimmer. 

A natural evolution in the field of active colloids is to consider chemically active filaments. The three-dimensional geometry and the separation of scales into long and thin directions unlocks extra degrees of freedom, functionality and dynamics. This includes localisation of surface flow patterns through the modulation of surface properties and precision-control of their 3D positioning through the centreline shape. Dynamic control of the centreline shape, perhaps using external stimuli, could then be used to individually control the motion and function of slender microbots \citep{MicrotransformersTom2018}.  

Modelling approaches have evolved to predict the benefits of creating slender phoretic microswimmers. 
Initial phoretic models addressed slender filaments with only straight centrelines, with spheroidal or arbitrary cross-section \citep{yariv2013electrophoretic, schnitzer2015osmotic, michelin2017geometric}. This was followed by the development of slender phoretic theory (SPT) by \cite{SPT2020}, derived using matched asymptotics from a boundary integral formulation. Independently, \cite{poehnl2021phoretic} derived a slender phoretic theory for curved rods from a more intuitive perspective of distributions of sources and dipoles. SPT, as developed in \cite{SPT2020,SPT_AnalyticalSolutions2022,katsamba2024slender}, calculates the induced slip velocities via evaluation of line integrals in the diffusion-dominated regime. The theory is asymptotically accurate for long and thin filaments, and is valid for general non-intersecting three-dimensional filament centrelines, both open-ended and looped. Further, it can accommodate varying circular cross-sectional radius with arbitrary chemical patterning, including  discrete jumps in chemical patterning \citep{KatsambaMontenegroJohnson2024_BookChapter}. Notably, numerical implementations of SPT \citep{SPT2020,katsamba2024slender} are orders of magnitude faster than boundary element methods \citep{montenegro2015regularised}, which typically involve a large number of surface integrals. 

This previous modelling of chemically propelled filaments has focused on rigid structures with pre-defined, fixed shapes. Realistically, phoretic filaments are not perfectly rigid, and the forces exerted on the body due to relative fluid motion and the chemically generated slip flows may deform the filament, changing its geometry and, hence, its swimming behaviour. A natural question is therefore: how do flexible phoretic filaments behave dynamically? 

Microscale flexible filaments in viscous flows are well-studied, due in part to their prevalence in many biological and artificial systems \citep{laugapowers2009,duRoure2019dynamics}. 
In biology, carpets of slender filaments, known as cilia, actively pump fluid over cycles driven by internal active forces \citep{ishikawa2024fluid}; sperm cells swim due to active waves sent down their long and thin tail-like appendages \citep{gaffney2011mammalian}; and bacteria rotate slender, helical flagella to propel themselves \citep{lauga2016bacterial}.
Meanwhile, flexible fibres are present in many industrial and household processes, such as textile fibres in clothing \citep{duprat2022moisture}, wood pulp fibres in the production of paper \citep{lundell2011fluid}, and microplastics that can pollute the environment \citep{dibenedetto2025fluid}.%

The dynamics of these flexible filaments in viscous flows are difficult to simulate since there is intrinsic feedback between the filament shape, the distribution of forces and the surrounding fluid.
Simulations are often computationally expensive because the underlying equations are notoriously numerically stiff, and there are often also accompanying constraints, such as inextensiblity, that must be satisfied throughout the evolution of the system \citep{duRoure2019dynamics}. 
A range of methods have been used to model and simulate the hydrodynamics of slender flexible filaments. Local slender body theories, such as resistive force theory \citep{Hancock1953,gray1955propulsion,cox1970motion,batchelor1970slender,lighthill1976flagellar}, have been applied to generate fast, efficient simulations of elastic filaments \citep{moreau2018asymptotic,walker2020efficient}, while non-local slender body theories \citep{keller1976slender,johnson1977slender,johnson1980improved,Koens_Lauga_2018} provide greater accuracy through more intensive simulations \citep{tornberg2004simulating}. 
Alternative approaches involve using singularity methods such as regularised Stokeslets \citep{cortez2001method,cortez2005method,hallmcnair2019efficient}, immersed boundary methods \citep{pozrikidis1992boundary,peskin2002immersed} and bead models \citep[][and references therein]{Delmotte2015general}. 
These models have been developed further to investigate the interaction of flexible fibres and the behaviour of passive and active suspensions \citep{tornberg2004simulating,schoeller2021methods}.

Elastic filaments demonstrate a wide range of interesting and exciting behaviours, depending on the strength of applied forces relative to the stiffness of the material. 
As summarised in the review of \cite{duRoure2019dynamics}, as the strength of external forcing or the flexibility of passive fibres in a flow increases, the dynamics progress from behaving as rigid rods, to undergoing buckling instabilities, folding through hook shapes, and contorting out-of-plane, depending on the prescribed flow and surrounding environment.

Active fibres have also been found to undergo similar dynamic transitions depending on the strength of activity relative to the fibre stiffness, with a particular focus on clamped filaments as models of ciliary beating or actin filaments. 
Bead-spring models of a  
chain of elastically linked beads, such as with stresslets distributed along its axis \citep{laskar2013hydrodynamic} or when forced at one end \citep{laskar2017filament}, demonstrated whirling, corkscrewing and planar beating dynamics; 
similar linear, helical and planar modes were found when attached to a passive cargo \citep{manna2017colloidal}.  This bead-spring approach has recently been extended to finer resolution to investigate active poroelastic filaments under `follower' forces that act tangentially to the centreline at the free end of the filament \citep{altunkeyik2025dynamics}, which showed rich transitions between dynamic behaviours, including for chemically active poroelastic filaments, and exhibited regimes of periodic and diffusive-like motion.

An alternative approach to studying active elastohydrodynamics is to combine resistive force theory with the mechanics of elastic filaments. 
When the motion is confined to lie in the plane,
filaments deforming due to follower forces have been shown to undergo Hopf bifurcations, seen as a flapping motion for clamped filaments \citep{decanio2017spontaneous},
while a two-dimensional linear stability analysis revealed transitions to rotation or beating that depends on mechanical boundary conditions, applied load, and noise effects \citep{fily2020buckling}. 
Further dynamical regimes have been explored using this elastohydrodynamic theory when applying localized tangential forces \citep{man2019morphological} to extract motions such as straight swimming, circling, spiralling and undulating, and also with clamped filaments in confined domains \citep{stein2021swirling}, revealing transitions of filaments with distributed tangential forces from stable to oscillatory and streaming modes.
Recent theoretical contributions by \cite{clarke2024bifurcations} and  \cite{schnitzer2025onset} further investigate the full three-dimensional motion of these follower-force models, identifying beating and whirling onset via bifurcation theory and weakly nonlinear analysis, enriching the fundamental understanding of these dynamics. 
Interestingly, clamped filaments that are free to deform in three-dimensions display a non-planar `whirling' motion following the onset of instability, and only reach a planar beating state when the applied force is much stronger.
In a related system, \cite{lough2023self} modelled the swarmer cell state of \emph{Proteus mirabilis} as an active Kirchoff rod, and found different swimming modes that may be planar or twisted into three-dimensions, depending on the relative material stiffnesses.  
Although many of these models abstract away the details provided by the surface chemistry of phoretic propulsion, for example by using follower forces, we expect that they retain essential features of force-induced shape transformations and the resulting complex locomotion, and highlight the complex interplay of elasticity, active forcing, and boundary conditions in shaping filament dynamics.

In this paper, we combine recent advances in the efficient simulation of elastohydrodynamic filaments \citep{walker2020efficient} with the asymptotically-derived framework for calculating slip flows in chemically active, diffusiophoretic filaments \citep{SPT2020,SPT_AnalyticalSolutions2022,katsamba2024slender} to investigate the fundamental dynamic behaviours of individual chemically propelled elastic filaments. We consider examples of canonical surface chemical patterning to study the transition between qualitatively different dynamic regimes as the material stiffness is varied. Section \ref{sec:Theory} outlines our theoretical method and numerical implementation for modelling chemically active filaments, using a local slender body theory to calculate the resulting slip flows from chemical reactions on the filament's surface, which are applied as a boundary condition in the numerical simulations of the elastohydrodynamics.
We present our results in section \ref{sec:Results}, investigating typical behaviours as the material stiffness (or equivalently the activity) is varied, via two key examples: a symmetrically-patterned filament that buckles under its own induced slip flow, and an asymmetrically-patterned filament whose swimming trajectory is altered by its deformation. We discuss and analyse the outcomes further in section \ref{sec:Discussion}.

\section{Theory of chemoelastohydrodynamic filaments} \label{sec:Theory}

\begin{figure}
    \centering
    \includegraphics[width=0.55\textwidth,trim = 1cm 2cm 0cm 2cm,clip]{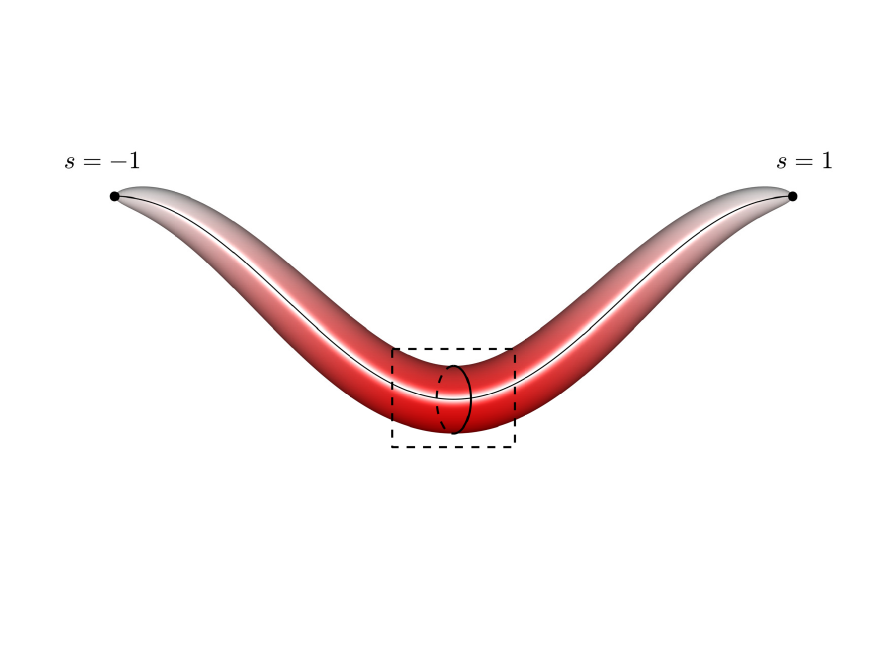}
    \includegraphics[width=0.35\textwidth,trim = 1cm 0cm 1cm 0cm,clip]{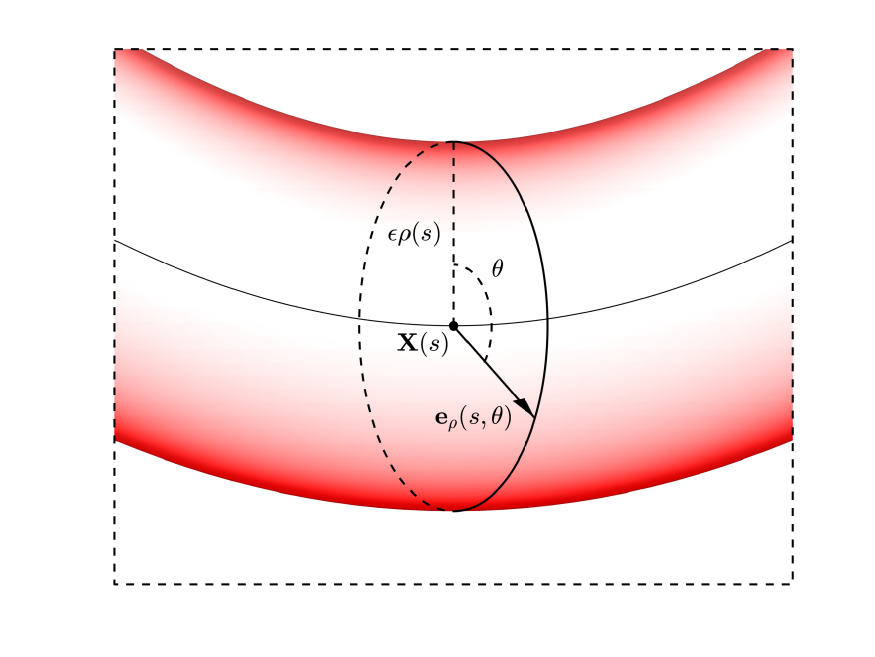}
    \caption{A slender chemically active elastic filament in an infinite fluid bath. The filament geometry is captured by a centreline, $\Xc(s)$, parametrised by its arclength $s$, with local polar coordinates on any circular cross-section given by $(\rho,\theta)$. Chemical patterning can vary across the surface of the filament, as shown by the filament colour, and generates solute at rate $\activity(s,\theta)$. Any resulting surface solute concentration gradients generate a slip flow that moves the surrounding fluid and applies viscous forces on the filament that can propel and deform it. 
    }
    \label{fig:schematic}
\end{figure}

We consider an inextensible slender elastic filament in an otherwise quiescent, infinite viscous fluid, as shown in \cref{fig:schematic}. 
Within the fluid is a chemical solute that diffuses rapidly. 
The solute acts as a fuel for the filament's self-propulsion: a chemical reaction of the solute dispersed in the surrounding fluid is catalysed by the catalytic coating on the surface of the filament, and short-range interactions between solute molecules and the filament's surface generate propulsive slip flows from solute concentration gradients within a thin layer by the filament's surface.
The resulting bulk flow can propel and deform the filament, altering the filament conformation and changing the solute distribution;
there is a complex feedback between shape, chemical patterning and swimming trajectory that we wish to explore.

Here, we outline the main theoretical components used to capture the evolution of slender chemically propelled elastic filaments. Our work combines two pre-existing methods: slender phoretic theory \citep{SPT2020,SPT_AnalyticalSolutions2022,katsamba2024slender, KatsambaMontenegroJohnson2024_BookChapter} that asymptotically calculates the leading order slip flow around slender chemically active filaments when given their chemical patterning and shape, and an efficient method for simulating elastohydrodynamics \citep{walker2020efficient}. Together, these allow us to probe the dynamic behaviour of chemically propelled elastic filaments, which we refer to as chemoelastohydrodynamics. 

\subsection{The slender filament geometry}

The geometry of a slender filament with a circular cross-section is captured by its centreline position, $\Xc(s,t)$, and cross-sectional radius $\crossradius(s)$, each parametrised by the arclength $s$ such that $|\partial\Xc/\partial s|=1$. The filament centreline has a total length $2\length$ and the maximum cross-sectional radius is $\maxcrossradius$, and we assume the filament is unshearable, so that cross-sections remains perpendicular to the centreline. We define a slenderness parameter by the ratio of the maximum diameter to the total arclength, $\epsslend=\maxcrossradius/\length$, which can be considered an aspect ratio, and we assume that our filaments are long and thin so that $\epsslend\ll 1$.

The surface of the filament is patterned with
two important surface chemical properties: the activity, $\activity(\x)$, which prescribes a surface solute flux (either generating or depleting solute), and the mobility, $\mobility(\x)$, that quantifies how chemical gradients drive slip flows.
Theoretically, we can set these independently and they may vary over the surface, but we assume the distribution remains fixed in the material frame.

The system is non-dimensionalised: spatial coordinates are rescaled by the half-length of the filament $\length$, while the cross-sectional radius is rescaled by the maximum radius $\maxcrossradius$. Taking typical (maximum) values for the activity and mobility as $\activityscale$ and $\mobilityscale$, respectively, we also use a concentration scale $\activityscale\maxcrossradius/D$ and a velocity scale $\activityscale\mobilityscale\maxcrossradius/ D \length$ to rescale our equations, where $D$ is the diffusivity of the solute. All equations are henceforth given in dimensionless form. 

\subsection{Local slender phoretic theory} \label{sec:localSPT}

The slip flow generated by the chemical activity is calculated using slender phoretic theory. The full details of this asymptotically-accurate theory are given in \citet{SPT2020,SPT_AnalyticalSolutions2022,katsamba2024slender,KatsambaMontenegroJohnson2024_BookChapter}; here we outline the key points.

We consider the zero P\'{e}clet number limit for the solute dynamics, where diffusion dominates advection, so that the solute concentration, $c$, in the surrounding fluid is governed by Laplace's equation,
\begin{equation}
    \nabla^2 c = 0, \label{eq:LaplaceEqn}
\end{equation}
where we consider the disturbance from a background concentration, so that $c\to 0$ far from the surface. 

The activity provides a normal flux boundary condition at the filament surface via zeroth order reaction kinetics, and any resulting  surface concentration gradients drive the slip flow proportional to the mobility. These are boundary conditions on the surface of the filament
\begin{align}   
    &-\no\cdot\boldsymbol{\nabla} c = \activity(\bv{x}), \label{eq:activity}
    \\
    \vslip &= \mobility(\x) \left(\idmat - \no \no \right)\cdot\boldsymbol{\nabla} c, \label{eq:mobility}
\end{align}
where $\no$ is the outward normal to the solid surface. 
Importantly, we frame our results in terms of the arclength $s$ and local polar coordinates in the cross-section $(r,\theta)$. Surface properties are therefore defined by the arclength $s\in[-1,1]$ along the centreline and angle $\theta\in[-\pi,\pi]$ around the cross-section, i.e.~$\activity=\activity(s,\theta)$ and $\mobility=\mobility(s,\theta)$, with the surface at $r=\rho(s)\in[0,1]$. Note that we take $\theta$ as the angle relative to a fixed position in each cross-section in the material frame. These local coordinates are shown in the inset to \cref{fig:schematic}.

In the framework of slender phoretic theory (SPT) \citep{SPT2020,  SPT_AnalyticalSolutions2022,  katsamba2024slender, KatsambaMontenegroJohnson2024_BookChapter}, the solution of Laplace's equation for the solute concentration is derived via a matched-asymptotics expansion in the slenderness, $\epsslend$, on the boundary integral solution of Laplace's equation.  This method systematically reduces the problem from solving an implicit indirect surface integral equation over the filament to a significantly simpler explicit calculation of a line integral along the filament centreline. This \emph{non-local} slender theory for chemically active filaments is valid for arbitrary non-self-intersecting three-dimensional centrelines, and can deal with both varying thickness and non-axisymmetric chemical patterning. The solution for the surface solute concentration for an axisymmetric activity is given in \cref{app:nonlocalSPT}.

In this work, we use a \emph{local} version of SPT, consistent with our use of a local slender body theory in the hydrodynamics discussed in \Cref{sec:RFT}. 
For this, the surface solute concentration is expanded as
\begin{align} \label{eq:conc_expansion}
    c(s,\theta) = \zerothorder{c}(s,\theta;\epsslend) + \epsslend \firstorder{c}(s,\theta;\epsslend) +
    o(\epsslend),
\end{align}
where the terms $\zerothorder{c},~\firstorder{c}$ are allowed to depend weakly (i.e.~logarithmically) upon $\epsslend$, and we calculate only the leading order component of each of these terms.This corresponds to neglecting the non-local line integral contributions in each asymptotically derived term of the full slender phoretic theory (quoted in \cref{app:nonlocalSPT}), since they are $O(1/\log\epsslend)$ smaller than the local terms. The resulting slip velocity over the surface of the swimmer can then be determined as a surface gradient of the concentration
 \begin{align}
    \frac{\vslip(s,\theta)}{\mobility(s,\theta)}  
    =& \underbrace{  \frac{\etheta(s,\theta)}{\epsslend \crossradius(s)} \frac{\partial \zerothorder{c}}{\partial \theta} }_{O(1/\epsslend)}
    +\underbrace{\left[  \frac{\etheta(s,\theta)}{ \crossradius} \frac{\partial \firstorder{c}}{\partial \theta} 
    + \tanhat(s) \frac{\partial \zerothorder{c}}{\partial s} \right]}_{O(1)} + 
    o(1), 
\label{eq:SlipVelocity}
\end{align}
where $\tanhat(s)$ is the tangent to the centreline and $\etheta(s,\theta)$ is the unit vector pointing azimuthally around the cross-section.
Note that gradients in concentration in the azimuthal direction around a cross-section, i.e. in the $\theta$ direction, are magnified by a factor $1/\epsslend$ in the slip velocity. 

Local slender phoretic theory calculates the leading order slip velocity, which is required to determine the leading order behaviour of the resulting dynamics.
However, it is important to note 
that reassembling the surface solute concentration from local SPT using \cref{eq:conc_expansion} is no longer asymptotically accurate, as we neglect some intermediate terms by removing the line integral contributions.
Crucially, due to the magnification of $\theta$-gradients of $\firstorder{c}$, 
the leading order slip velocity is accurately calculated, since we retain terms of $O(\log\epsslend)$ and ignore terms of $O(1)$ which are smaller. A comparison of local and non-local SPT is given in \Cref{app:local_vs_nonlocal}, showing that local SPT well-approximates the slip velocity.

\subsubsection*{Axisymmetric activity}

We focus on the case of an axisymmetric activity, $\activity=\activity(s)$. For this form of chemical patterning, in both local and non-local slender phoretic theories, the leading order surface solute concentration is also axisymmetric and, hence, independent of $\theta$: $\zerothorder{c}(s,\theta)=\zerothorder{c}(s)$ \citep{SPT2020}.
In fact, the leading order and next order concentrations of local SPT are found to be 
\begin{subequations}
    \begin{align}
        \zerothorder{c}(s) \label{AxisymmActivityConc_c0}
&=  
\frac{1}{2}
\crossradius(s)\activity(s)\log\left(\frac{4(1-s^2) }{\epsslend^2 \crossradius^2(s)}\right), \\
\firstorder{c}(s,\theta)
&=
     \frac{1}{2} \crossradius^2(s) \curvature(s) \activity(s)	
    \cos\thetamode \left[\log\left(\frac{4(1-s^2)}{\epsslend^2 \crossradius^2(s)}  \right)
    - 3\right],
    \end{align}
\end{subequations}
where $\kappa$ is the Frenet-Serret curvature and $\Theta$ is the azimuthal angle relative to the Frenet-Serret basis, tracking the azimuthal angle relative to the cumulative torsion (note that $\Theta$ is not fixed relative to the material frame and the Frenet-Serret normal coincides with $\Theta=0$).
We note that the leading order surface concentration is uniform over any cross-section, $\zerothorder{c}=\zerothorder{c}(s)$, and so the leading order slip velocity is $O(1)$. 

To determine the leading order kinematics of the filament, it is only necessary to calculate the cross-sectional average of the slip velocity, since all higher order modes in a Fourier series in $\theta$ have smaller effect in the motion of the slender body \citep[see, for example,][\S2.9]{SPT2020}.
The slip velocity calculation can then be further simplified by decomposing $\etheta$ into Frenet-Serret normal and binormal vectors, so that we find the cross-sectional average slip velocity to be
\begin{align} \label{eq:axi_avg_slip}
    \vslipo(s) \equiv \frac{1}{2\pi} \int_{-\pi}^{\pi} 
    \vslip(s,\thetadum)
    ~\dd \thetadum = \mobility(s)  
    \left[
    \tanhat(s) \frac{\partial \zerothorder{c}}{\partial s}
    -\frac{1}{2}\norhat(s)\coeffsinmode(s)  
    \right] 
\end{align}
where $\norhat(s)$ is the Frenet-Serret normal and 
\begin{align}
    \coeffsinmode(s)=& -
    \frac{1}{2}\left[\log\left(\frac{4(1-s^2)}{\epsslend^2 \crossradius^2(s)}\right) -3\right]\crossradius(s)\curvature(s)\activity(s).  \label{eq:coeffsinemode}  
\end{align}

\subsection{Resistive force theory} \label{sec:RFT}

The phoretic slip flow induces bulk flow in the surrounding fluid, acting as a boundary condition to the Stokes equations for a viscous fluid,
\begin{align*}
    \mu \nabla^2 \vec{u} = \nabla p,
    \qquad
    \nabla\cdot\vec{u} = 0,
\end{align*}
where $\mathbf{u}(\mathbf{x},t)$ is the fluid velocity, $p(\mathbf{x},t)$ is the fluid pressure and $\mu$ is the fluid's dynamic viscosity. The flow is assumed to decay to $\mathbf{u}=\vec{0}$ in the far field, and satisfies a slip boundary condition on the surface of the body.

To model the Stokes flow around the slender filament, we use resistive force theory \citep{Hancock1953,gray1955propulsion}. This is a local slender body theory for viscous flow that decomposes the local motion of the filament (relative to a background or slip flow) into components that are locally tangential and normal to the centreline, linearly relating the leading order force density and velocity of each via different drag coefficients. Here, the appropriate velocity is the surface velocity in the lab frame (the frame in which $\bv{u}\to0$ far from the filament), which is $\vec{u}_{motion} + \vslipo$, where $\vec{u}_{motion}$ corresponds to the local velocity of the slender body. Note that this captures the notion that slip velocities act in the opposite direction to the motion that they drive, so that a rigid body moves in the opposite direction to the mean slip velocity on its surface (in the absence of other effects). We decompose this combined quantity into $\uparallel$ and $\uperp$, the components in the tangential and normal directions, respectively. The leading order force density is linearly related to this local relative velocity via
\begin{align}
    \forcedens_\parallel = - C_\parallel \uparallel,
    \qquad
    \forcedens_\perp = - C_\perp \uperp,
\end{align}
where $\forcedens_\parallel, \forcedens_\perp$ denote the force density applied on the filament in directions tangential to the centreline and normal to it. The drag coefficients are
\begin{align}
    C_\parallel = \frac{2\pi\mu}{\log(4/\rho\epsslend)-0.5},
    \qquad
    C_\perp = \frac{4\pi\mu}{\log(4/\rho\epsslend)-0.5}.
\end{align}
The resistive torque theory of \citet{WalkerResistiveTorqueTheory} links local angular velocity (such as that about the local tangent) to local torque via an additional resistive coefficient. Following this, here we take the local applied torque per unit length to be $4\pi\epsilon^2\rho(s)^2\mu\bm{\Omega}$ for angular velocity $\bm{\Omega}$ of the slender body. This application of resistive force theory is valid for the slip flows generated by chemically-propelled filaments, since local slender body theories are consistent with boundary conditions that enforce rigid-body motion locally, as is the case here \citep{WalkerResistiveTorqueTheory}.

\subsection{Kirchoff rod equations}
\label{sec:elastohydrodynamics}

We model the deformation of the filament using the  Kirchhoff equations, which are the pointwise force and moment balance equations, 
\begin{align}
    &\frac{\partial \Finternal}{\partial s} + \forcedens = \bv{0},    \label{KirchhoffForceBalance}
\\
    \frac{\partial\Mbending}{\partial s} &+ \tanhat\wedge\Finternal + \torquedens =\bv{0}. \label{KirchhoffTorqueBalance}
\end{align} 
where $\Finternal$ is the  internal force acting on the cross-section of the rod at $s$, and $\Mbending$ is the bending moment. We identify $\forcedens$ and $\torquedens$ as the force and torque densities due to the viscous tractions generated by the flow; these are computed using the resistive force framework of \cref{sec:RFT}.

We impose conditions of zero force and torque at the filament ends, so that
\begin{align}
    \Finternal(-1) = \Finternal(1) = \Mbending(-1) = \Mbending(1) = 0.
\end{align}
Integrating \cref{KirchhoffForceBalance,KirchhoffTorqueBalance} leads to the governing equations
\begin{align}
    \int_{-1}^{1} &\forcedens(\sdum) ~\dd\sdum = 0,
    \\
    \int_{s}^{1} [\Xc(\sdum)-\Xc(s)]& \times \forcedens(\sdum) + \torquedens(\sdum) ~\dd\sdum = \Mbending(s).
\end{align}
The bending moment is given by the constitutive equation
\begin{align}
    \Mbending 
    = EI \left( 
    \kappa_1 \cdarboux{1}
    + \curvature_2 \cdarboux{2}
    + \frac{\curvature_3}{1+\nu} \cdarboux{3}
    \right),
\end{align}
where $\cdarboux{i}$ are a right-handed orthonormal director basis with $\cdarboux{3}=\tanhat$, and $\curvature_i$ are the components of the twist vector such that $\partial \cdarboux{i}/\partial s = \bv{\kappa}\times\cdarboux{i}$. The material parameters are captured by the bending stiffness, $EI$, and the Poisson ratio, $\nu$; for simplicity, we assume both of these are constant throughout the filament, with $\nu=1$ \citep{antman2005nonlinear,nizette1999towards}.

\subsection{Numerical implementation}
 
To solve this system numerically, we discretise the slender body into $N$ straight segments of equal length. We assign a local orthonormal director basis to each segment, fixed in the body so that it tracks the material deformation of the body. These director bases are parametrised using Euler angles, employing independent coordinate systems for each segment. Necessarily, this introduces singularities in the parametrisation that may be identified with the poles of canonical spherical polar coordinate systems and related to the gimbal lock phenomenon. These singularities are sidestepped numerically following \citeauthor{walker2020efficient}'s method: during the dynamics, when any of the Euler angles approaches the singularities of their respective coordinate system, a new coordinate system is chosen for each segment such that the poles are maximally distant from the current configuration. In effect, this adaptive reparametrisation enables us to utilise the convenience of an Euler-angle parametrisation whilst avoiding the singularities associated with this approach.

We direct the interested reader to the work of \citet{walker2020efficient} for a full and detailed account of their  approach. Omitting full definitions for brevity, the dimensionless discretised chemoelastohydrodynamic system takes the form
\begin{equation}
   \elastohydronum BAQ\dot{\vec{\Theta}} = - \vec{R},
\end{equation}
where the vector $\vec{\Theta}$ records the locally defined Euler angles and the position of one end of the slender body, from which the current configuration can be uniquely determined. The linear operators $B$, $A$, and $Q$ encode various aspects of the theory: $Q$ converts from the angular parameterisation to the lab-frame coordinates of the endpoints of the $N$ discrete segments; $A$ uses the resistive force theory of \citet{Hancock1953,gray1955propulsion} and the resistive torque theory of \citet{WalkerResistiveTorqueTheory} to relate linear and angular velocities to the forces and torques on the slender body; $B$ encodes the governing equations of force and torque balance on the slender body. Accordingly, the vector $\vec{R}$ records any applied forces and torques on the body. In this case, $\vec{R}$ includes elastic restoring moments and the forces and torques caused by the phoretic slip velocity.

The key dimensionless parameter is the elastohydrodynamic number,
\begin{equation}
    \elastohydronum = \frac{8\pi\mu (2\length)^4}{EI\,T} ,
    \label{eq:EH_number}
\end{equation}
which is defined in terms of the fluid viscosity $\mu$, the half-length $\length$ of the slender body, the bending stiffness $EI$ of the body, and a characteristic timescale $T$. Here, we take $T=D\length^2/\activityscale\mobilityscale\maxcrossradius$ as the characteristic timescale set by slender phoretic theory. The elastohydrodynamic number therefore represents the ratio of the elasto-viscous timescale, $t_{ev} \propto\mu \length^4/EI$, to the flow timescale $T$. It increases with chemical activity and mobility, and decreases with increasing filament bending stiffness, so that high $\elastohydronum$ corresponds to strong slip flows or very flexible filaments.

The linear system defined by $BAQ$ is square, so that one may readily solve for $\dot{\vec{\Theta}}$ to yield evolution equations for the position and configuration of the slender body. These are implemented in MATLAB and solved using the adaptive, implicit timestepping scheme \texttt{ode15s} \citep{Shampine1997}. In adapting the implementation of \citet{walker2020efficient}, we are implicitly assuming that the slender body is inextensible, unshearable and moving in a Newtonian fluid at vanishing Reynolds number..

\section{Results} \label{sec:Results}

\begin{figure}
    \centering
    \begin{tikzpicture}
        \node[anchor = west,font = \large] at (0,0) {(a)};
        \node[anchor = west,font = \large] at (4.2,0) {(b)};
        \node[anchor = west,font = \large] at (9.8,0) {(c)};
        \node[anchor = west] at (15,0) {};
    \end{tikzpicture}
    \begin{tikzpicture}
        \node at (0,-0.5) {\includegraphics[width=4cm,trim = 4cm 8cm 4cm 8cm,clip]{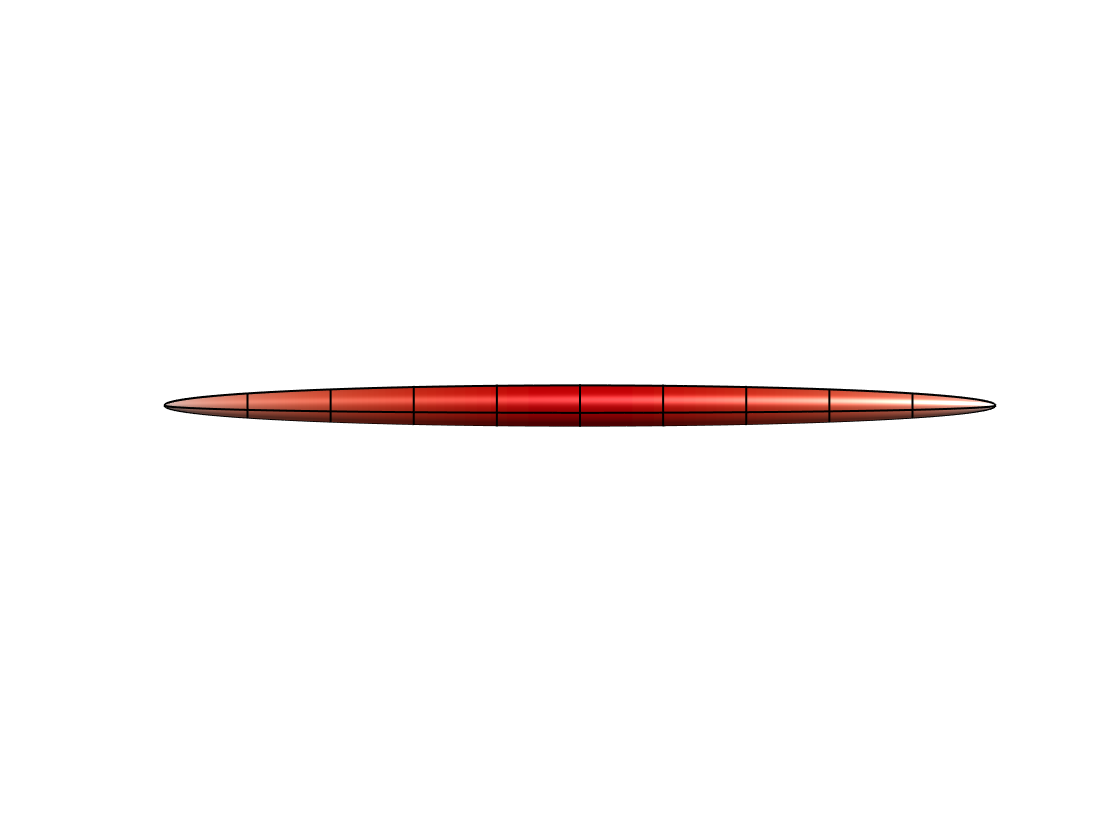}};
        \node at (0,0) {$\activity = \sqrt{1-s^2}$};

        \node at (0,-2.5) {\includegraphics[width=4cm,trim = 4cm 8cm 4cm 8cm,clip]{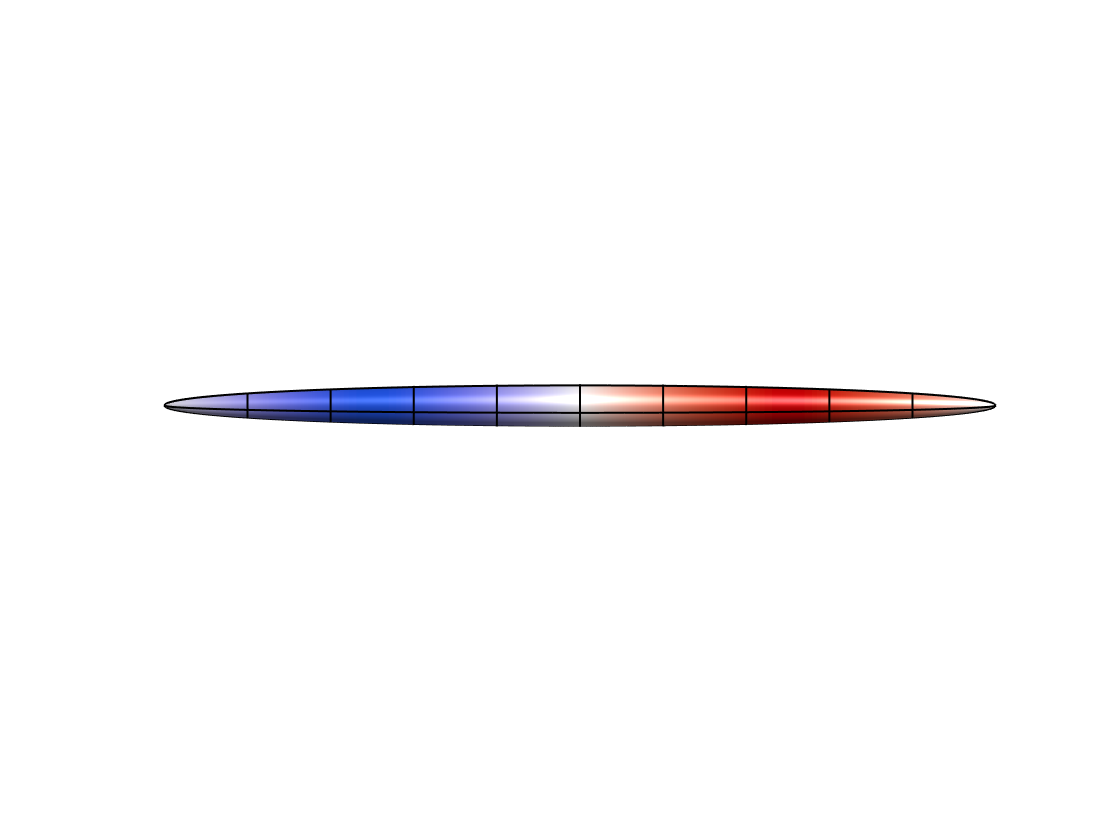}};
        \node at (0,-2) {$\activity = \sin(\pi s)$};
        \node at (8,-1.75) {\includegraphics[width=11cm]{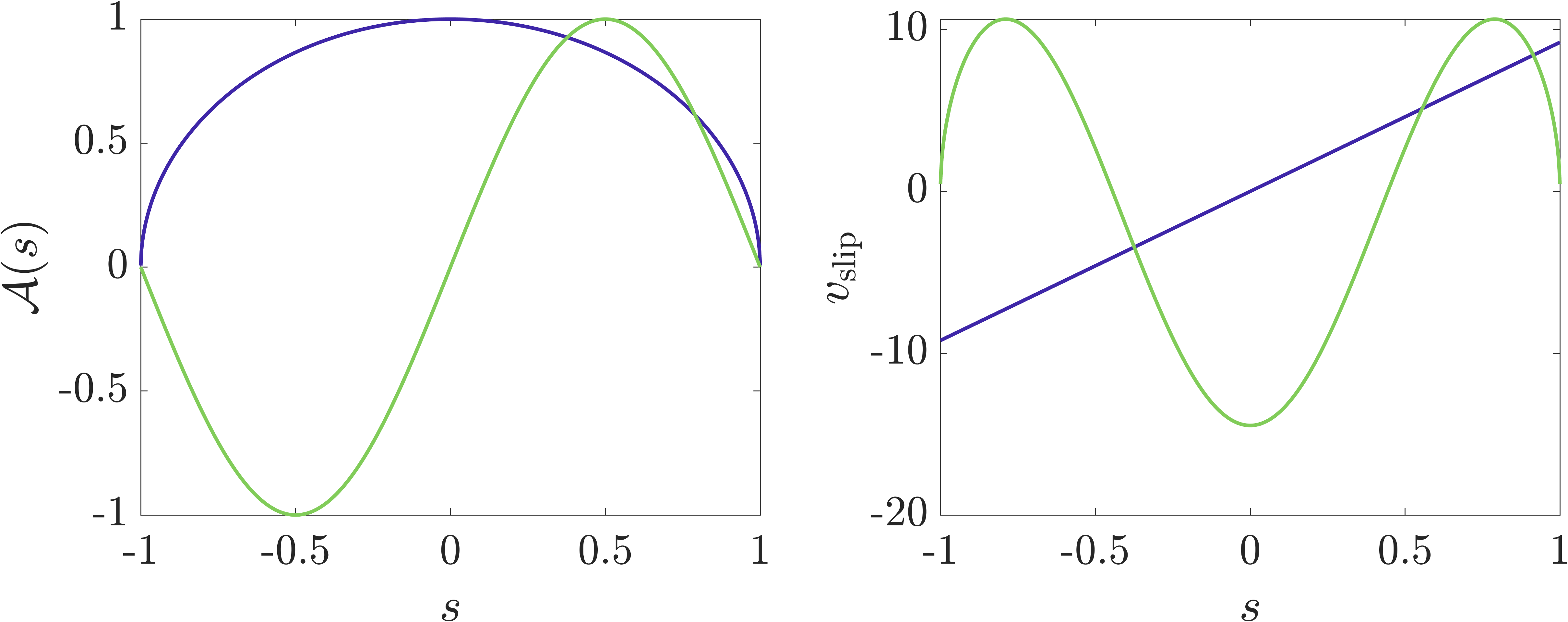}};
        \matrix [draw=none] at (7,-2.4) {
        \node[blue!60!black,font=\tiny] {$\activity = \sqrt{1-s^2}$}; \\
        \node[green!60!black,font=\tiny] {$\activity = \sin(\pi s)$}; \\ 
        };
        \matrix [draw=none] at (12.6,-2.4) {
        \node[blue!60!black,font=\tiny] {$\activity = \sqrt{1-s^2}$}; \\
        \node[green!60!black,font=\tiny] {$\activity = \sin(\pi s)$}; \\ 
        };
    \end{tikzpicture}

    \caption{The chemoelastic filaments considered. All shapes are prolate spheroidal, with cross-sectional radius $\crossradius = \sqrt{1-s^2}$. 
    The considered activities are $\activity = \sqrt{1-s^2}$ and  
    $\activity = \sin(\pi s)$.
    (a) Representations of the activity and shape of the filaments in their elastic rest state. 
    (b) The prescribed activity 
    and (c) the slip velocity calculated from local SPT for the square root (blue) 
    and sinusoidal (green) activities, when $\mobility=-1$. 
    The symmetrically-patterned filament generates a surrounding extensional flow that applies compressive forces on the filament, while the antisymmetric filament has regions of both compression and extension. 
    } 
    \label{fig:Activities}
\end{figure}

In this section, we present an anthology of results for the intricate  dynamics of chemoelastohydrodynamic filaments, exploring long-time, three-dimensional simulations for different chemical activity profiles. We focus on a prolate spheroidal geometry, $\crossradius(s)=\sqrt{1-s^2}$, and activity profiles that decay to zero at the filament tips. Our theory is valid for more general profiles and chemical patterning than this \citep[see e.g.~][]{SPT2020}; however, we focus on these examples to avoid more technical hurdles that might arise in choosing an appropriate discretisation to account for boundary layer effects \citep{SPT_AnalyticalSolutions2022} or from the breakdown of the slenderness assumptions of SPT near the ends, $s = \pm 1$, for shapes that taper too rapidly. We thus consider two key axisymmetric activity profiles, 
\begin{equation}
    \text{(a) } ~\mathcal{A} = \sqrt{1 - s^2}, 
    \hspace{1cm}
    \text{(b) } ~\mathcal{A} = \sin{\pi s},
\end{equation}
which represent (a) a symmetric smooth activity profile, 
and (b) an antisymmetric smooth activity  profile. In the rigid analogue, these profiles would behave as (a) Saturn and (b) Janus rods, respectively. Unless otherwise stated, the initial configuration is an S-shaped planar curve, which is deformed from the elastic rest state of a perfectly straight filament. The activity and tangential slip velocity for these filaments in the straight rod configuration is shown in \Cref{fig:Activities} for a mobility $\mobility=-1$. Note that, when straight, the symmetric activity (a) generates an extensional flow in the surrounding fluid, from the centre to the tips; this causes compression in the elastic filament as the phoretic surface is propelled against the slip flow. For the antisymmetric example (b), there are regions of both compression and extension.

We focus on exploring the effect that varying the elastohydrodynamic number $\elastohydronum$ has on the swimming dynamics, in a similar manner to studies based upon follower forces \citep[e.g.][]{laskar2017filament,decanio2017spontaneous,fily2020buckling,man2019morphological,clarke2024bifurcations, schnitzer2025onset,altunkeyik2025dynamics} and force distributions \citep[e.g.][]{laskar2013hydrodynamic,manna2017colloidal,stein2021swirling,lough2023self}.  We vary $\elastohydronum$ within the range $\elastohydronum = 1000$ (very stiff, low activity) to $\elastohydronum = 100,000$ (very deformable, high activity), looking for characteristic behaviours. In particular, we are interested in examining the transitions between different dynamic regimes; for instance, the initial buckling of an elastic filament, and transitions between planar, periodic and chaotic behaviours that may arise as the filament becomes more deformable (or, equivalently, as the active forces increase). 

\subsection{Saturn-like activity: Fore-aft symmetric profiles} \label{sec:SqrtActivity}

A perfectly rigid, straight rod with an activity $\activity=\sqrt{1-s^2}$ does not move and generates a straining flow where fluid is pushed outwards towards the poles of the rod. We therefore expect the filament to be under a compressive loading and, thus, exhibit buckling for sufficiently high values of the elastohydrodynamic number $\elastohydronum$. Similar phenomena are observed for filaments driven by non-conservative follower forces, as discussed in the introduction.

However, in contrast to follower force models, not only do we have tangential forces (potentially both extensional and compressive) distributed along the entire length of the filament, but also azimuthal slip flows, which at leading order generate normal forces on the filament in the plane of curvature. These normal forces may serve to suppress or enhance the buckling behaviour locally, depending on the sign of the mobility, and so offer some potentially rich additional behaviours. 

We consider the dynamics of these filaments, beginning with stiff filaments that have low values of $\elastohydronum\approx1000$, and increasing to $\elastohydronum=100,000$. This increase in $\elastohydronum$ can be considered as increasing the deformability (decreasing stiffness) with fixed activity, or equivalently increasing the activity with fixed stiffness. 

\subsubsection{Planar buckling: low $\elastohydronum$}

For very low values of the elastohydrodynamic number, $\elastohydronum$, the viscous forces due to the activity-induced flows are weak compared to the elastic forces. The filaments quickly straighten from any deformed perturbation, and little motion is observed. 

\begin{figure}
    \centering
     \begin{tikzpicture}
        \node[anchor = west,font = \large] at (0,0) {(a)};
        \node[anchor = west,font = \large] at (4.2,0) {(b)};
        \node[anchor = west,font = \large] at (9.5,0) {(c)};
        \node[anchor = west] at (15,0) {};
    \end{tikzpicture}
    \includegraphics[width=0.9\linewidth]{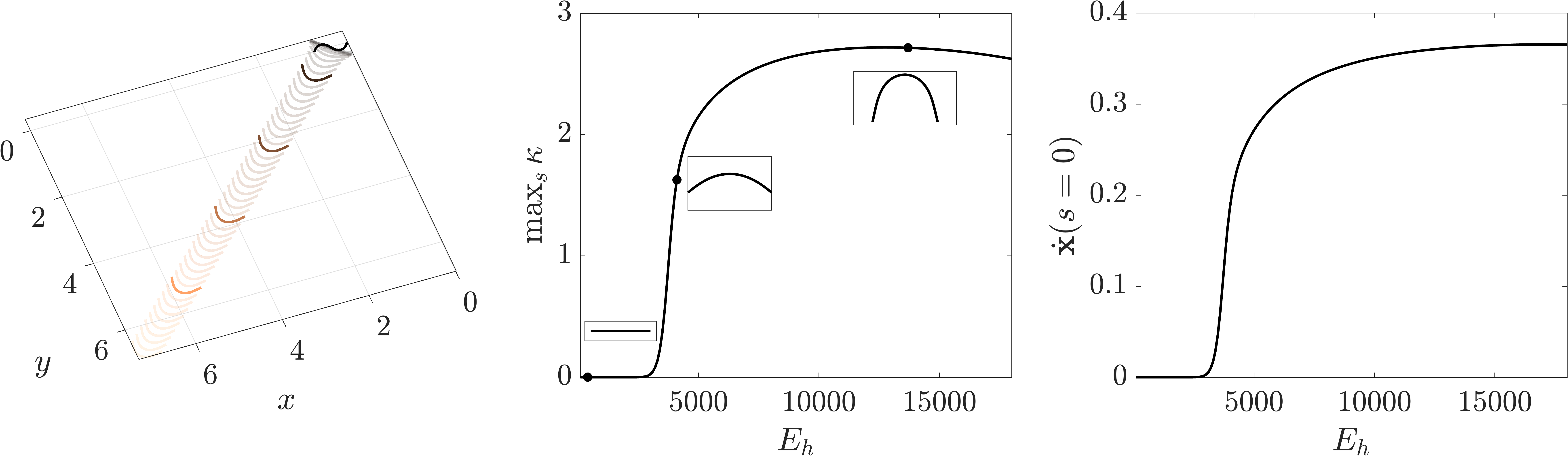}
    \caption{Planar ballistic motion of a filament with symmetric activity $\activity=\sqrt{1-s^2}$ at low $\elastohydronum$. (a) An example trajectory with $\elastohydronum=5000$ up to $t=40$, with time progressing from dark to light. 
 (b) The maximum curvature at steady state as a function of $\elastohydronum$, with insets showing example filament shapes, computed until $t=10$. (c) The ballistic speed increases with $\elastohydronum$ before plateauing. 
    }
    \label{fig: pump to translator}
\end{figure}

Once $\elastohydronum$ is above a critical value, close to $\elastohydronum\approx2000$, we observe a transition in behaviour:
the filament deforms to a clear U-shape, and moves forwards on a ballistic, planar trajectory away from the tips of the U-shape. An example trajectory is shown in \Cref{fig: pump to translator}a for $\elastohydronum=5000$. 
 This transition is reminiscent of Euler buckling; starting from a small perturbation from a straight rod, the compressive forcing due to the chemical activity is large enough to buckle the filament into its first Euler mode, and the resulting shape asymmetry generates concentration gradients between the inside and outside of the curve that drive swimming. This buckling instability unlocks new modes of swimming via shape change  from I-shaped pumps to U-shaped swimmers \citep{sharan2021fundamental,MicrotransformersTom2018}. 

In this U-shaped swimming, we observe a settling towards a steady propelling state, which varies with $\elastohydronum$. For example, as $\elastohydronum$ increases, so does the curvature of the filament at steady state, as shown in \Cref{fig: pump to translator}b. This can be explained, for example, by considering the same compressive load being applied to a relatively more flexible filament, promoting further bending. With both ends of the filament pointing more directly backwards, more propulsive force is aligned with the direction of motion; the filament must remain force-free and so it moves more quickly to balance this force via viscous drag, which can be seen in \Cref{fig: pump to translator}c. Long-time simulations indicate that these planar trajectories are likely stable for elastohydrodynamic numbers as large as $\elastohydronum \approx 21,000$. Even from out-of-plane helical initial conditions, planar behaviour seems to be a stable attractor for $\elastohydronum$ up to approximately $20,000$.

\subsubsection{Onset of three-dimensional motion: moderate $\elastohydronum$}

As the elastohydrodynamic number increases further, we observe a transition from planar motion to fully three-dimensional dynamics, with the filament leaving the initial plane of motion. Beginning with our initial planar S-shape perturbation, we observe this clearly within time $t=100$ from $\elastohydronum \approx 21,500$.

\begin{figure}
    \centering
     \begin{tikzpicture}
        \node[anchor = west,font = \large] at (0,0) {(a)};
        \node[anchor = west,font = \large] at (4.5,0) {(b)};
        \node[anchor = west,font = \large] at (9.6,0) {(c)};
        \node[anchor = west] at (15,0) {};
    \end{tikzpicture}
    \includegraphics[width=0.9\linewidth]{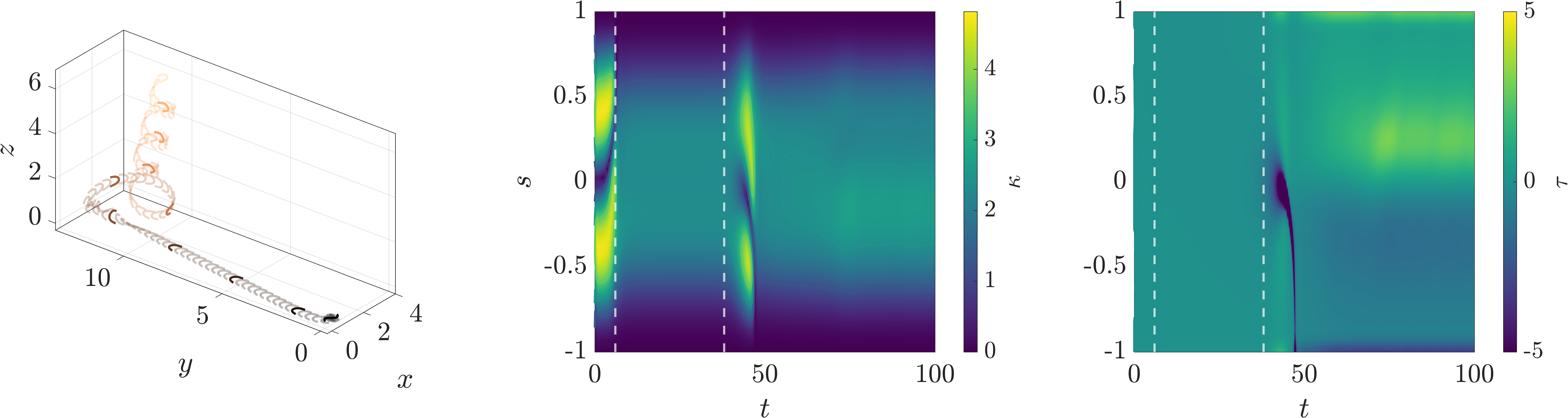}
    \caption{At long times and above a threshold of $\elastohydronum\approx21,500$, trajectories deviate out-of-plane. Here, an example is shown for $\elastohydronum=23,000$. (a) Trajectory snapshots up to time $t=100$, with time progressing from dark to light, showing the initially planar trajectory begin to spin, and then sharply turn to settle into a near-helical trajectory. 
    Heatmaps of (b) the filament curvature and (c) torsion demonstrate a clear transition in geometry and dynamics. 
    }
    \label{fig: trans to rolling trans}
\end{figure}

Just above this threshold, the initially planar filament adopts a U-shaped translating mode for a period of time, but then small amounts of out-of plane motion cause it to begin revolving about its direction of motion as it progresses forwards.
As $\elastohydronum$ increases further, above approximately $23,000$, the out-of-plane spinning increases in angular frequency, the filament begins to wobble and, for higher values of $\elastohydronum$, we observe a sharp, non-planar turn followed by coherent ballistic helical motion out of the plane. This latter mode comprises a translational component and a spin about the axis of translation, with a small wobble. Extended simulations suggest that the resulting `wobbly helix' trajectory is stable when it exists. This behaviour is shown for $\elastohydronum = 23,000$ in \Cref{fig: trans to rolling trans}.

\subsubsection{Stable periodic orbits: high $\elastohydronum$} \label{sec:pinwheeling}

\begin{figure}
    \centering
     \begin{tikzpicture}
        \node[anchor = west,font = \large] at (0,0) {(a)};
        \node[anchor = west,font = \large] at (4.5,0) {(b)};
        \node[anchor = west,font = \large] at (9.6,0) {(c)};
        \node[anchor = west] at (15,0) {};
    \end{tikzpicture}
    \includegraphics[width=0.9\linewidth]{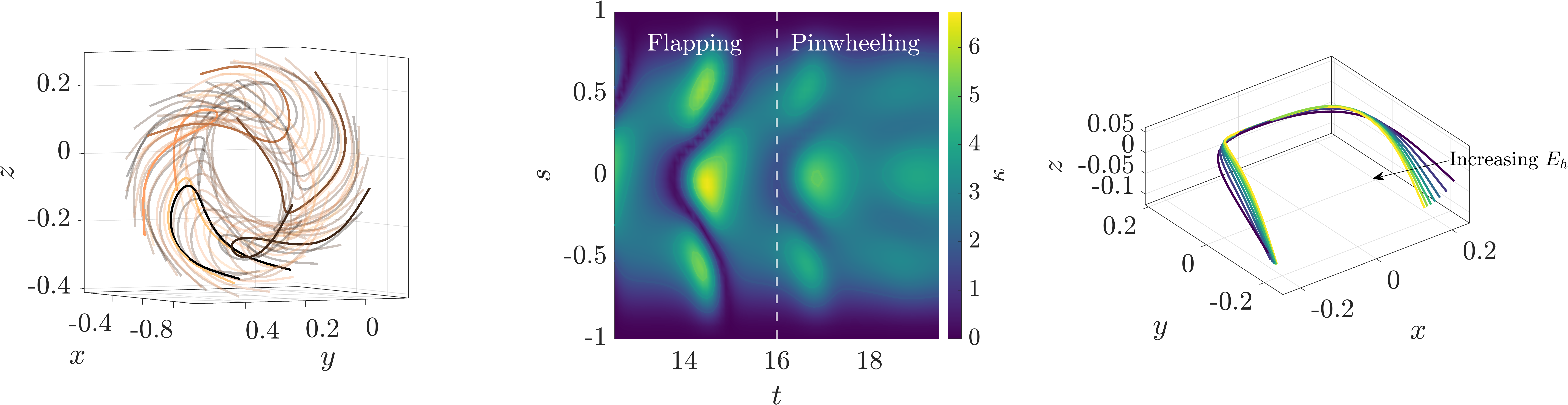}
    \caption{Stable periodic orbits exist at intermediate $\elastohydronum$. (a) Snapshots of a stable `pinwheeling' trajectory, with time progressing from dark to light, showing its characteristic tight circular motion for $\elastohydronum = 32,000$. 
    (b) A heatmap of the curvature shows an initial period of symmetric periodic flapping motion that gives way to the stable fixed shape of pinwheeling. 
    (c) While orbiting, the filament is in an approximately fixed configuration, that can be described as a folded or bent over U-shape. As $\elastohydronum$ increases, the U-shape becomes increasingly non-planar, resulting in tighter circular motion. }
    \label{fig: pinwheeling}
\end{figure}

The out-of-plane helical trajectories described above undergo a critical transition to an apparently periodic state as $\elastohydronum$ increases slightly further. For example, after initial transients, the tight helical path of $\elastohydronum = 23,000$ is lost by $\elastohydronum=24,000$, whereafter the early-time planar translation deviates quickly to a tight, circular path that does not translate over long timescales. An example of this is shown in \Cref{fig: pinwheeling} 
for $\elastohydronum=32,000$. We refer to this state as `pinwheeling'.

This pinwheeling behaviour, once exhibited, seems to be a stable attractor. As $\elastohydronum$ increases, the early-time dynamics prior to pinwheeling goes through a range of ballistic trajectories, including planar translation, flapping back and forth and more chaotic-like motion, but these
trajectories soon transition to the stable orbiting motion. These orbits appear to exist up to $\elastohydronum \approx 48,000$.

During pinwheeling, the shape of the filament appears to be approximately fixed; these shapes are illustrated in \Cref{fig: pinwheeling}c. The filament is symmetric about the midpoint and takes the form of a U-shape that is bent out-of-plane near the tips. This bend becomes more pronounced as $\elastohydronum$ increases, resulting in a tighter pinwheeling trajectory, i.e.~motion on a small circular attractor.

\subsubsection{Diffusive-like motion: very high $\elastohydronum$} 

\begin{figure}
    \centering
     \begin{tikzpicture}
        \node[anchor = west,font = \large] at (0,0) {(a)};
        \node[anchor = west,font = \large] at (5,0) {(b)};
        \node[anchor = west,font = \large] at (10,0) {(c)};
        \node[anchor = west] at (15,0) {};
    \end{tikzpicture}
    \includegraphics[width=0.9\linewidth]{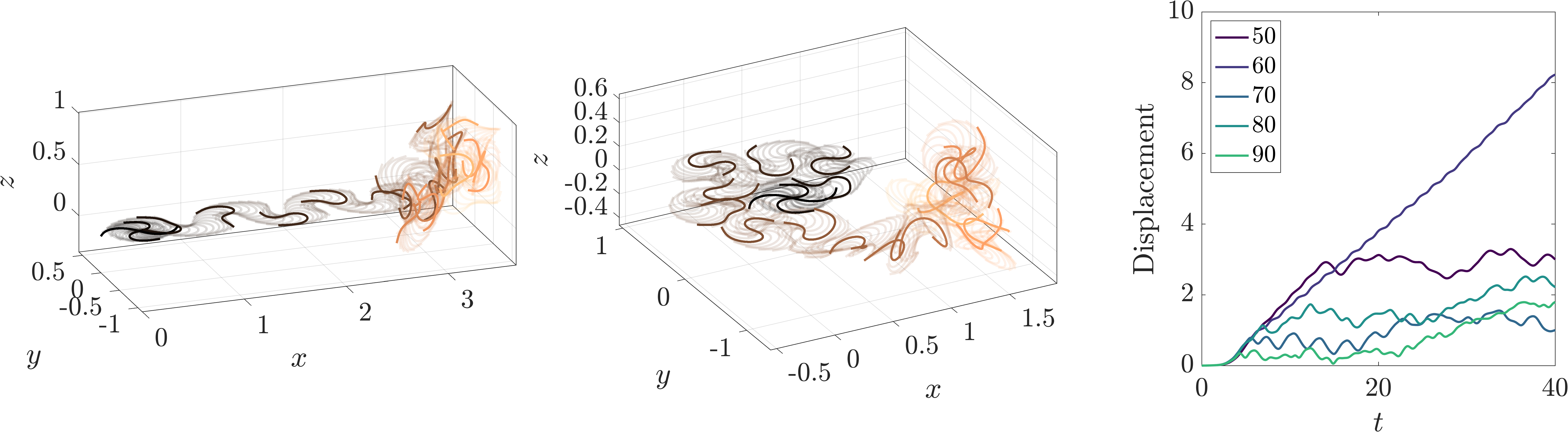}
    \caption{Diffusion-like behaviour at high $\elastohydronum$. Snapshots of the trajectory of filament with (a) $\elastohydronum = 50,000$ up to time $t=40$, and (b) $\elastohydronum=70,000$ up to a time $t=40$. In each, time progresses from dark to light. After a short period of initial planar motion, the filament undergoes regular tumbling and reorientation in a chaotic, diffusion-like manner.
    (c) Filament displacement for various elastohydrodynamic numbers, with the legend denoting $\elastohydronum/1000$. As $\elastohydronum$ increases, we can see the progression from trapped periodic orbits, to more persistent directed motion, before reaching diffusive-like dynamics at very large $\elastohydronum$.}
    \label{fig: high EH dynamics}
\end{figure}

The pinwheeling trajectory, however, cannot become tighter indefinitely. Once $\elastohydronum \approx 49,000$, early-time transient dynamics gives way to complex, three-dimensional motion. The active forces are sufficiently strong compared to the elastic forces that the filament deforms significantly, resulting in sharp changes in direction and speed. This behaviour looks qualitatively like a random walk, and perhaps even chaotic. Example trajectories are shown in \Cref{fig: high EH dynamics}a,b.

The progression in behaviour of the chemoelastic filaments can be clearly characterised by considering the midpoint displacement from its initial position over time, as shown in \Cref{fig: high EH dynamics}c. The presence of significant ballistic-like motility can be seen clearly for $\elastohydronum=60,000$, with dynamics that are more similar to diffusion or random walks occurring for the other values shown.

\subsubsection{Persistent ballistic motion} 

\begin{figure}
    \centering    
     \begin{tikzpicture}
        \node[anchor = west,font = \large] at (0,0) {(a)};
        \node[anchor = west,font = \large] at (4.7,0) {(b)};
        \node[anchor = west,font = \large] at (9.8,0) {(c)};
        \node[anchor = west] at (15,0) {};
    \end{tikzpicture}
    \includegraphics[width=0.9\linewidth]{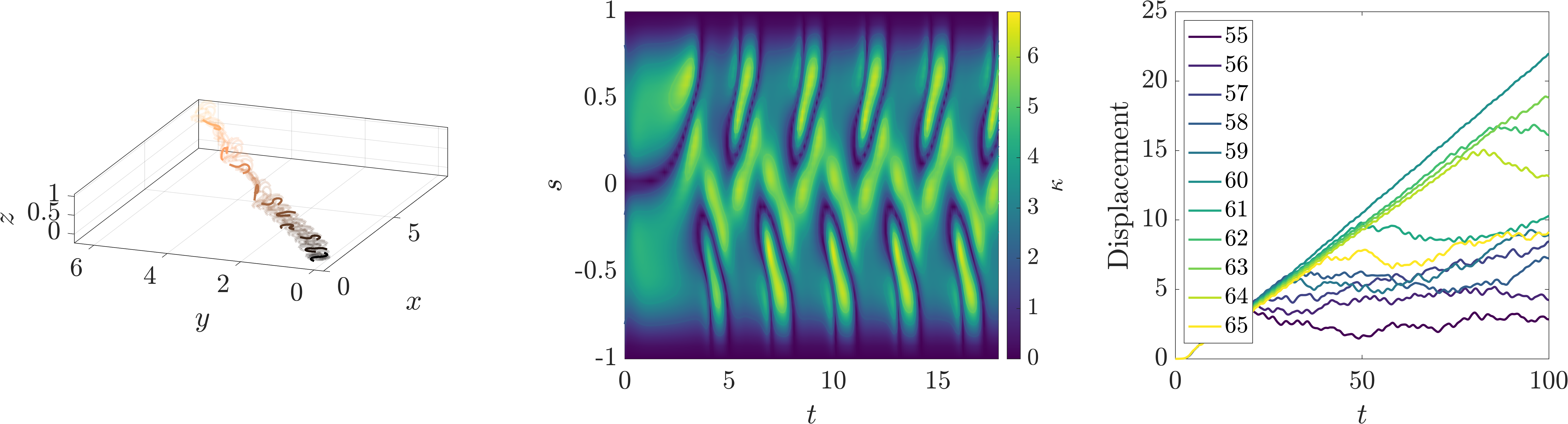}
    \caption{Within a band of high values of $\elastohydronum$, a ballistic-like motion appears to persist.
    (a) Trajectory for $\elastohydronum=60,000$ for time $t=100$, with time progressing from dark to light. 
    (b) Curvature heatmap for the same simulation, showing waves of curvature travelling outwards from the centre of the filament. (c) The displacement over time for a range of values of $\elastohydronum$, showing (planar) motion that persists for intermediate $\elastohydronum$.
    }
    \label{fig: stable flappers}
\end{figure}

Prompted by the large displacement seem in \Cref{fig: high EH dynamics}c, we focus on the dynamics for values near $\elastohydronum=60,000$. as shown in \Cref{fig: stable flappers}c. This reveals more examples of apparently meta-stable planar dynamics, characterised by sustained `flapping' motion. An example trajectory is shown in \Cref{fig: stable flappers}a for $\elastohydronum=60,000$.
Additional simulations from a helical perturbation suggest that these dynamics can be reached from non-planar states, and so may be a stable (or metastable) attractor for some set of parameter values.

The progressive flapping motion comprises a travelling wave of curvature, that initiates in the middle of the filament, and alternates its direction of propagation towards either end of the filament, as seen in \Cref{fig: stable flappers}b. As such, progress is a rocking back and forth motion along a nearly straight trajectory. 

\subsection{Janus-like activity: Fore-aft antisymmetric profile} \label{sec:SinActivity}

Qualitatively different behaviours are observed for `Janus'-style rods, where the activity profile is antisymmetric about the half-way cross-section. With this chemical patterning, a perfectly straight and rigid filament would translate tangential to its axis due to the self-generated concentration gradient between the front and back, contrasting the stationary pumping flow of the symmetric `Saturn' examples previously considered. 

As the elastohydrodynamic number, $\elastohydronum$, is increased, we see transitions away from this swimming behaviour, with the filament buckling due to the viscous stresses from the active flow, and the resulting shape change causing circling and spiralling motions. Here, we consider an activity $\activity=\sin\pi s$, which is fore-aft antisymmetric and has extrema in activity a quarter of the way from the ends, at arclengths $s=\pm1/2$.

\subsubsection{Planar translation and circling: low $\elastohydronum$}

\begin{figure}
    \centering
  \begin{tikzpicture}
        \node[anchor = west,font = \large] at (0,0) {(a)};
        \node[anchor = west,font = \large] at (7.5,0) {(b)};
        \node[anchor = west] at (15,0) {};
    \end{tikzpicture}
    \begin{tikzpicture}
        \node at (-7,0) {\includegraphics[width=0.4\textwidth]{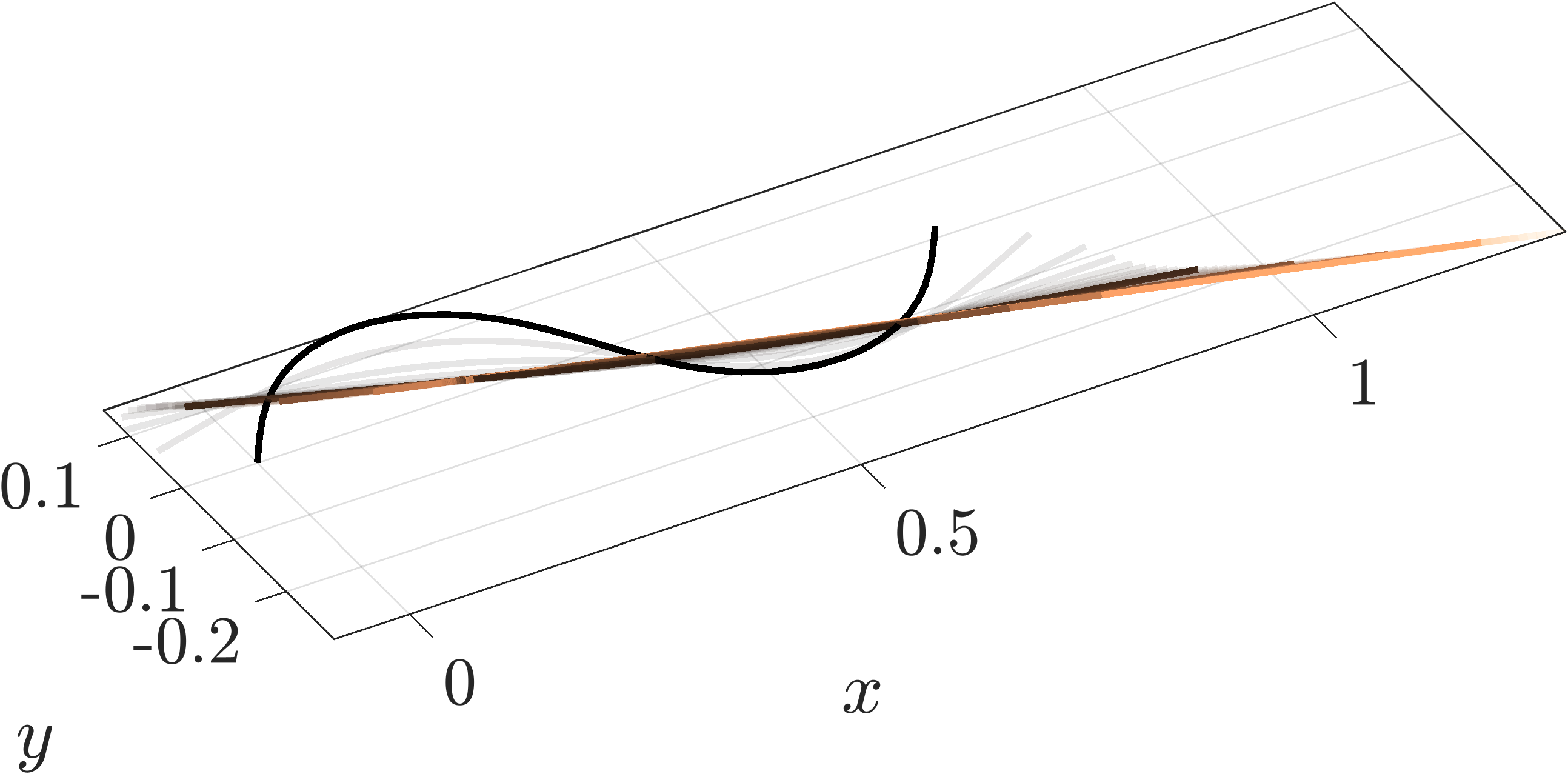}};
        \node at (0,0) {\includegraphics[width=0.4\textwidth,trim = 4cm 8cm 4cm 8cm,clip]{results_figs/activities/OptionB_sineactivity_diverging.png}};
        \node[text width = 2cm] at (1.6,1) {\footnotesize High $c$};
        \node[text width = 2cm] at (0,1) {\footnotesize Low $c$};

        \draw[->,-stealth, line width=1.25pt] (0.75,0.5)  -- (-0.25,0.5);
        \draw[->,-stealth, line width=1.25pt] (-2,0.5)  -- (-1,0.5);
        \draw[->,-stealth, line width=1.25pt] (1.5,0.5)  -- (2.5,0.5);

        \draw[->,-stealth, line width=1.25pt] (0.75,-0.25)  -- (-0.25,-0.25);
        \draw[->,-stealth, line width=1.25pt] (-2,-0.25)  -- (-1,-0.25);
        \draw[->,-stealth, line width=1.25pt] (1.5,-0.25)  -- (2.5,-0.25);
    \end{tikzpicture}
    \caption{A stiff, antisymmetrically-patterned filament, with $\activity = \sin \pi s$, straightens and propels along its axis at low $\elastohydronum$. (a) The evolution of a filament for $\elastohydronum=2000$ up to time $t=2$, with time progressing from dark to light. 
    (b) The straight filament has slip flows (denoted by arrows) directed towards the minimum in activity, and away from the maximum. The filament experiences viscous forces opposing the slip flow, resulting in the lower activity region towards the rear experiencing extension, while the higher activity region at the front is under compression. If the active forcing is sufficiently high compared to the stiffness, then the region under compression may buckle.
    }
    \label{fig:straight_swim}
\end{figure}

When the filament is sufficiently stiff, elastic restoring forces dominate the dynamics, causing any shape perturbation to decay. Hence, the filament relaxes back to its straight configuration, propelling itself along its axis. We observe this behaviour for elastohydrodynamic numbers approximately lower than $5000$, with an example shown in \Cref{fig:straight_swim}a.

\begin{figure}
    \centering
    \begin{tikzpicture}
        \node[anchor = west,font = \large] at (0,0) {(a)};
        \node[anchor = west,font = \large] at (4.2,0) {(b)};
        \node[anchor = west,font = \large] at (9.2,0) {(c)};
        \node[anchor = west] at (15,0) {};
    \end{tikzpicture}
    \includegraphics[width=0.9\linewidth]{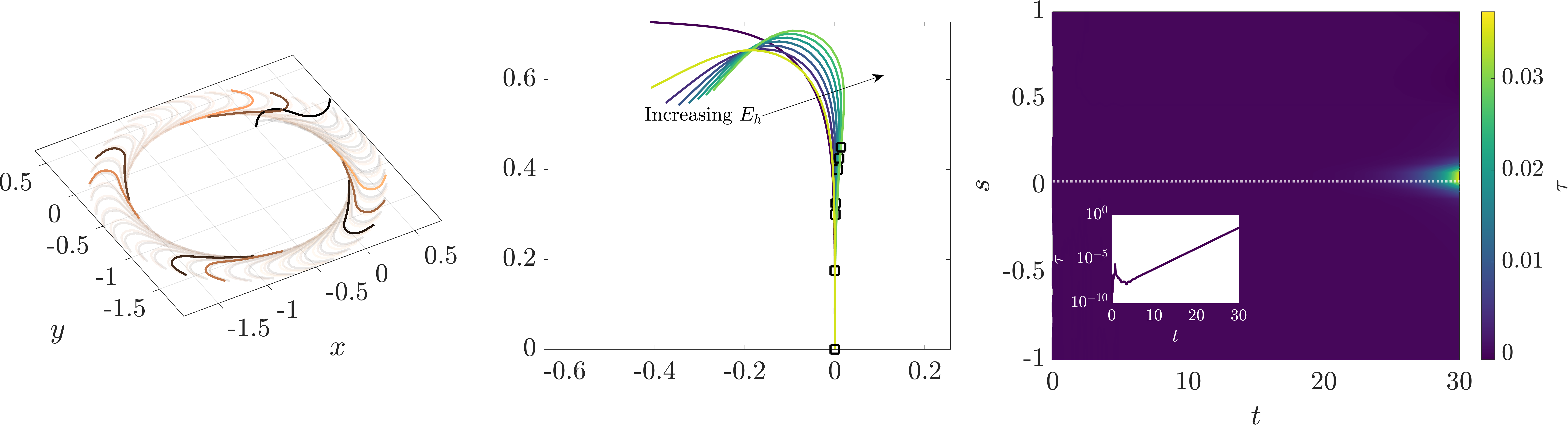}
    \caption{Buckling of the filament results in a planar circling motion. (a) Example circling trajectory for $\elastohydronum=28,000$ for a time $t=30$ with time progressing from dark to light. 
(b) Steady state hook shapes for $\elastohydronum=6,000$--$28,000$. Squares show the location of maximum torsion. (c) Torsion heatmap for $\elastohydronum=28,000$ showing a localised increase in torsion towards the filament centre. Inset: at a fixed point along the filament (given by the white line), the torsion increases exponentially over time. 
    }
    \label{fig:circling}
\end{figure}

The concentration gradients generated by the antisymmetric activity, $\activity=\sin\pi s$, result in regions of both compression and extension in the filament, due to the slip flows illustrated in \Cref{fig:straight_swim}b. The regions under compression are susceptible to buckling when the activity causes a strong enough flow relative to the material stiffness.
Indeed, as $\elastohydronum$ increases further, beyond $\elastohydronum=5000$, the viscous forces acting on the self-propelled filament are sufficient to bend it. In particular, we observe deformation into a hook-shaped bend at the front end of the filament, with the corner of the hook centred close to the most active region (see \Cref{fig:circling}b). 

Once in this configuration, there is a left-right asymmetry both in the distribution of the solute concentration (and the associated slip velocity) due to the activity sink on the longer leg of the hook, as well as in the resulting drag on each of the legs due to their different lengths. The outcome is that the hook shape propels at an angle to the line that bisects the corner of the hook, resulting in an overall circling motion. An example of this type of motion is shown in \Cref{fig:circling}a 
for $\elastohydronum=28,000$. As $\elastohydronum$ increases, the hook shape becomes more curved, and the trajectories follow a tighter circle.

\subsubsection{Transition to helical motion: moderate $\elastohydronum$}

\begin{figure}
    \centering
    \begin{tikzpicture}
        \node[anchor = west,font = \large] at (1,0) {(a)};
        \node[anchor = west,font = \large] at (8.5,0) {(b)};
        \node[anchor = west] at (15,0) {};
    \end{tikzpicture}
    \includegraphics[width=0.8\linewidth]{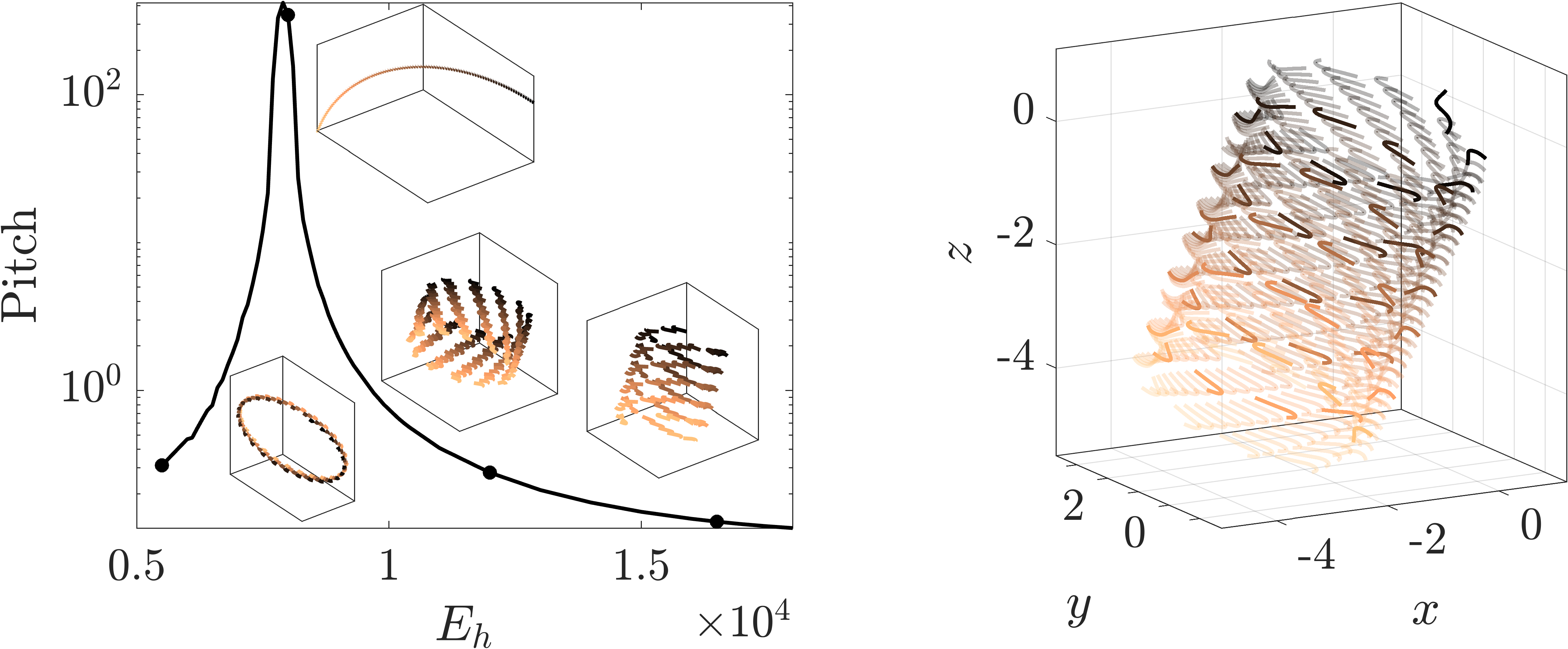}
    \caption{Filaments transition to three-dimensional motion via helical trajectories that, up to a point, become more pronounced with increasing $\elastohydronum$. (a) The helical pitch of trajectories as a function of $E_h$, with insets showing their motion. Pitch is highly non-monotonic: a sharp maximum is visible near $\elastohydronum\approx8,000$, with trajectories reducing their pitch and radius (not shown) after the peak. For the largest values of $\elastohydronum$ shown on this plot, the motion increasingly resembles quasi-planar dynamics, with slow out-of-plane drift. (b) Snapshots of a well-developed helical trajectory for $E_h=18,000$, with time progressing from dark to light. These simulations were initiated with the filament in a helical shape. }
    \label{fig: trans to helical swimmers}
\end{figure}

Similarly to the symmetric activity profiles, we observe a transition from planar motion to full three-dimensional dynamics at sufficiently high $\elastohydronum$. Here, this transition occurs around $\elastohydronum=5500$, so that planar circling appears to only be stable for a small range of values. Commensurate with this, in this regime the planar circling behaviour appears to be metastable, as our initial planar perturbation settles into a fixed hook-shape circling behaviour on relatively short timescales, but a localised growth in torsion leads to out-of-plane motion on longer timescales. An example of the growth in torsion is shown in \Cref{fig:circling}c for $\elastohydronum=28,000$, demonstrating the exponential local growth of torsion. This localised torsion, typically observed between the midpoint of the filament and the most-curved region of the hook (positions highlighted in \Cref{fig:circling}b), destabilises the planar motion of the filament, twisting the filament out-of-plane into fully three-dimensional trajectories. Simulating the same motion for longer times reveals a seemingly robust helical trajectory.

To fully understand which values of the elastohydrodynamic number permit helical trajectories, we simulate the long-time dynamics of our chemoelastic filament when starting from an initially three-dimensional helical perturbation, and fit a helix to the resulting midpoint trajectory.
In contrast to the sudden transition at a critical $\elastohydronum$ for the symmetrically-patterned filaments shown in \Cref{fig: pump to translator}, this out-of-plane transition appears gradually as $\elastohydronum$ increases. The resulting pitch of this helix is plotted as a function of $\elastohydronum$ in \Cref{fig: trans to helical swimmers}a, showing that pitch initially grows with $\elastohydronum$. As $\elastohydronum$ increases, the helix becomes significantly more pronounced and emerges over lengthscales many times that of the filament. Curiously, after a peak around $\elastohydronum\approx8,000$, the pitch decreases, leading to large, approximately circular trajectories that slowly move in three dimensions and trace out very large helices. Notably, these dynamics might easily be mistaken for planar motion on shorter timescales. An example of these dynamics, associated with significantly smaller scale helices than those before the peak in pitch, is shown in \Cref{fig: trans to helical swimmers}b.  We remark that the emergent radii of the helical trajectories follow the same trend as the measured pitches, such that the pitch serves as an approximate measure of the lengthscale of the helices.

\begin{figure}
    \centering
    \begin{tikzpicture}
        \node[anchor = west,font = \large] at (0,0) {(a)};
        \node[anchor = west,font = \large] at (4.6,0) {(b)};
        \node[anchor = west,font = \large] at (9.6,0) {(c)};
        \node[anchor = west] at (15,0) {};
    \end{tikzpicture}
     \includegraphics[width=0.9\linewidth]{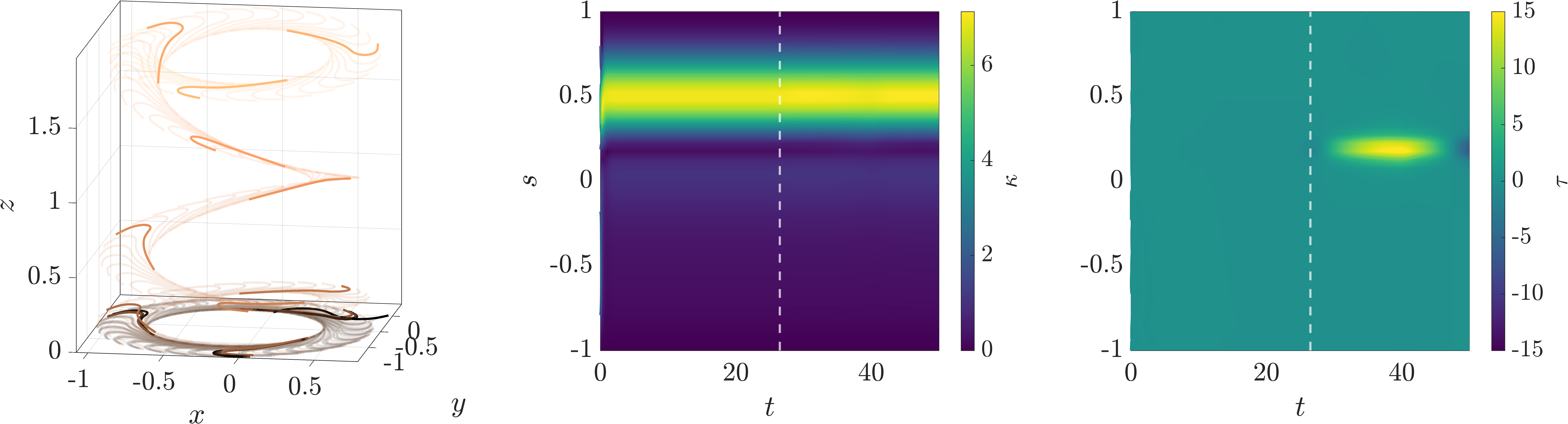}
    \caption{The transition from planar to three-dimensional motion. Here, we see helical motion for $\elastohydronum=50,000$ up to time $t=50$. (a) The trajectory showing an initial planar circling, before a sharp transition towards a helical spiralling motion out-of-plane, with time progressing from dark to light. 
Heatmaps of (b) curvature and (c) torsion, with only the latter showing a clear transition from planar to non-planar motion.}
    \label{fig:helical_spinning}
\end{figure}

These observations broadly align with simulations from an initially planar state, with helical motion of various forms present from around $\elastohydronum\approx5,500$. This suggests that this helical behaviour is a stable attractor of the dynamics, though a full and proper treatment of the solutions to the steady state dynamics is warranted to formally address the notion of stability. A typical example of this form of helical motion from the planar initial configuration is shown in \Cref{fig:helical_spinning}.
The large value of $\elastohydronum$ chosen here yields tight circles and a corresponding tight helix.
The motion initially remains planar for some time before transitioning to a helical path. At larger $\elastohydronum$, the out-of-plane motion appears to occur sooner and the resulting helix pitch increases. During the helical motion, the base state of the filament is a hook shape with a localised twist out-of-plane. As visible in \Cref{fig:helical_spinning}b,c, the curvature of the filament remains approximately unchanged throughout the motion (after initial transients), whilst the torsion undergoes a significant change that corresponds to the filament leaving the plane.

\subsubsection{Run-and-tumble and beyond: high $\elastohydronum$} 

\begin{figure}
    \centering
    \begin{tikzpicture}
        \node[anchor = west,font = \large] at (0,0) {(a)};
        \node[anchor = west,font = \large] at (7,0) {(b)};
        \node[anchor = west] at (15,0) {};
    \end{tikzpicture}
    \includegraphics[width=0.8\linewidth]{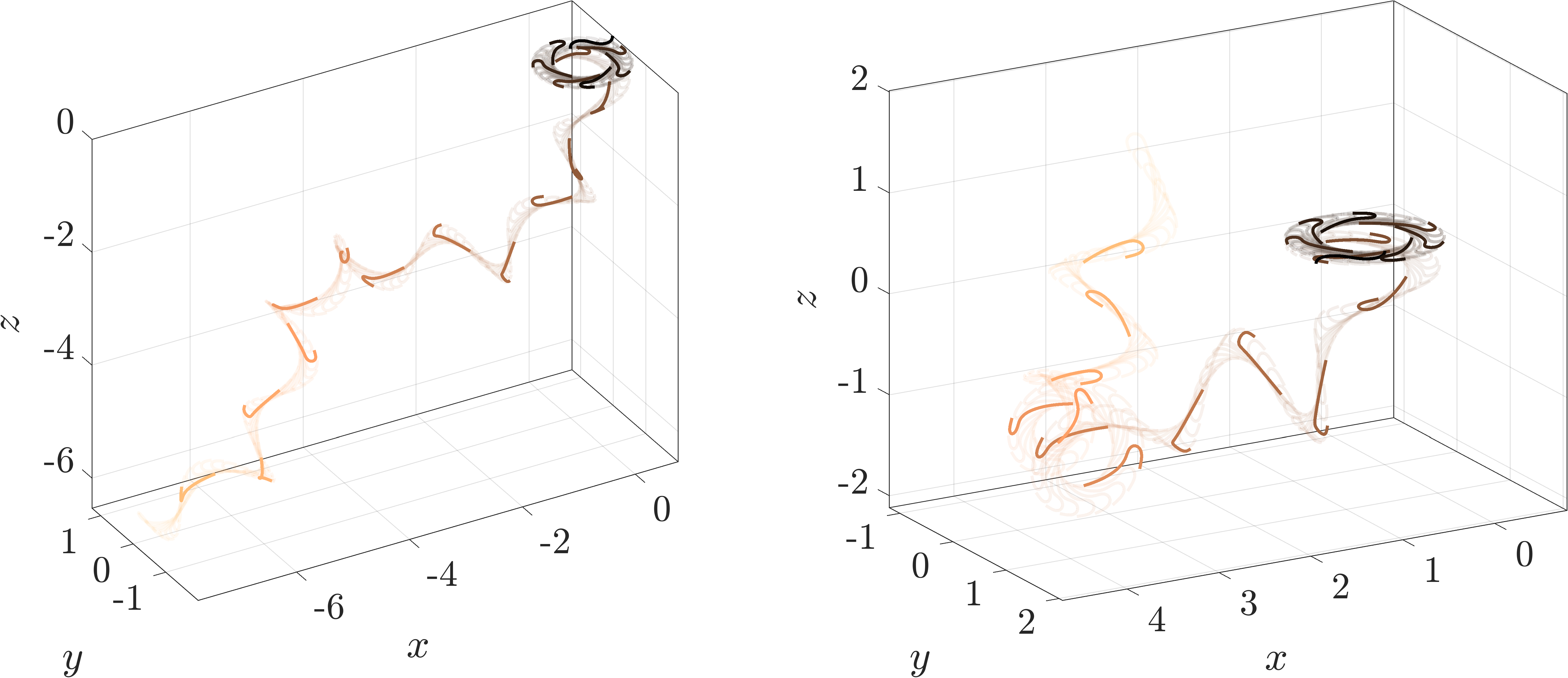}
    \caption{Beyond helical motion. Example trajectories for (a)  $\elastohydronum=80,000$ and (b) $90,000$, each up to time $t=50$, with time progressing from dark to light. The dynamics deviate from helical trajectories at large $\elastohydronum$, showing something akin to run-and-tumble motion at high $\elastohydronum$. 
}
    \label{fig:beyond_helix}
\end{figure}

When $\elastohydronum$ moves beyond approximately $80,000$, the helical dynamics are lost; the filament becomes more flexible, leading to more chaotic-looking paths, beginning with irregular, tightly wound helices. Examples of these trajectories are given in \Cref{fig:beyond_helix}.
The resulting trajectories are reminiscent of a persistent random walk, or perhaps the run-and-tumble dynamics of bacteria, with intervals of relatively persistent trajectories connected by sharp turns. 

This run-and-tumble-like behaviour may be explained by looking more closely at the filament shape during the dynamics. As the filament propels forward, the front end rapidly buckles from side-to-side in a chaotic manner, while the rear portion of the filament remains relatively straight. This is due to the slip flow in the front half applying locally compressional forces, leading to buckling, while the rear half experiences a locally extensional forcing, straightening the filament out. Whereas at lower $\elastohydronum$ the filament was both being pulled forward by the front half and pushed by the rear, here the filament may be thought of as mostly being pushed from the back half, while the rapid oscillations of the front half cause small side-to-side deviations. These deviations may cause small, seemingly random perturbations in the path direction, making the trajectory more like a persistent random walk.

\subsection{Variations}

In addition to the dynamics of the symmetric and antisymmetric examples considered above, we investigated some natural variations to more fully understand the type and sequence of dynamical behaviours observed in chemoelastic filaments. We outline a few key examples below.

\subsubsection{A near constant, fore-aft symmetric activity profile} \label{sec:TanhActivity}

To investigate whether the dynamical behaviours observed are specific to the examples considered, or if they give general behaviours for other activity profiles, we repeated our simulations with different activity profiles. Here we focus on the similarities and differences determined between the results of \Cref{sec:SqrtActivity} and a near-constant activity profile $\mathcal{A} = \tanh[10(1 - s^2)]/\tanh{10}$.

In general, the qualitative nature of the dynamics is largely unchanged from \Cref{sec:SqrtActivity}. We see the same dynamic regimes and transitions as $\elastohydronum$ increases: buckling from straight to planar U-shaped translation, transition to out-of-plane via twisting, trapped periodic pinwheeling orbits, persistent planar trajectories and diffusive-like motion. 

However, the values for $\elastohydronum$ where the behaviour transitions are noticeably higher. For instance, we first observe the stable periodic orbiting behaviours from $\elastohydronum \approx 30,000$, in contrast to the $24,000$ for $\activity=\sqrt{1-s^2}$. This is perhaps because, for this near-constant activity, the slip flows are weaker over the majority of the filament since concentration gradients are reduced, and so either a higher propulsive force or a weaker bending rigidity is required to cause the appropriate buckling transitions.

\subsubsection{Reversing the activity and mobility}

In the examples considered so far, we have focused on chemical patterning with positive activity and negative mobility, i.e.~the solute is generated at the filament surface and slip flows are directed down the surface concentration gradients. This choice resulted in straining flows that cause buckling in the symmetric filaments. A natural question to ask is: what happens if the activity and/or mobility change sign? Changing the sign of the activity causes the chemical patterning to deplete the solute, instead of generating it, and changing the sign of mobility causes slip flows to be directed up surface concentration gradients, rather than up them. 

Due to the linearity of our system of equations, reversing the sign of both the activity and mobility results in the same slip flows and, hence, identical dynamics. In other words, the sign of $\activity\mobility$ drives the dynamics. This means that chemical patterning which both depletes the solute, $\activity<0$, and causes slip flows up the chemical gradients, $\mobility>0$, have the same dynamics as those considered above. It is therefore only necessary to additionally consider what happens when $\activity>0$ and $\mobility>0$, which we consider for the chemical patterns from \Cref{sec:SqrtActivity,sec:SinActivity}.

For the symmetrically-distributed chemical patterning, $\activity=\sqrt{1-s^2}$, the flow is reversed compared to that in \Cref{sec:SqrtActivity}, leading to extensional viscous forces on the filament. Our simulations for $\elastohydronum<100,000$ all straighten on a fast timescale and no non-trivial elastohydrodynamics is observed.

Meanwhile, for antisymmetric chemical patternings, there is no observed difference in dynamics when reversing the sign of $\activity\mobility$. Indeed, reversing the sign of only one of the activity or mobility is equivalent to swapping the front and back of the filament, and so yields unchanged dynamics.

\subsubsection{Turning off the azimuthal component}

A key aspect of the theory of chemically propelled filaments that is uncommon in other physical systems is that the azimuthal slip flow (acting around the slender cross-sections, perpendicular to the local tangent) is as important to the leading order behaviour as the flow along the length of the filament, as can be seen in e.g.~\cref{eq:SlipVelocity,eq:axi_avg_slip}. It is not clear from our previous simulations what effect this component of the slip flow has on the observed dynamics.

To understand the importance of the azimuthal component of the flows, caused by the contribution of the $\firstorder{c}$ component of the slip flow in slender phoretic theory, as given in \cref{eq:SlipVelocity}, we ran simulations with this component artificially set to zero.
Qualitatively, we observe the same transitions in the dynamic behaviour for $\activity=\sqrt{1-s^2}$. As $\elastohydronum$ increases, the dynamics progress from straightening to planar U-shape translation, followed by an out-of-plane transition, then pinwheeling orbits, and eventually paths that appear more random and diffusive. However, these transitions occurred at significantly lower $\elastohydronum$ than the full simulations including the azimuthal component. For example, the initial buckling occurs below $\elastohydronum=1000$, with pinwheeling by $\elastohydronum=7000$. 

This perhaps suggests that the azimuthal component is stabilising the flow around the filament and resisting deformation to the next dynamic mode in the full-physics simulations. A possible explanation for this is that the slip flow suppresses buckling. If the filament begins to curve, then the positive activity generates relatively higher solute concentration on the inside of the curve than outside; for the negative mobility considered, the slip flows are then directed from the inside of the bend to the outside, propelling the bend in the opposite direction, back towards a straight configuration \citep{makanga2025instability}.
When these azimuthal components are removed, there are only tangential forces promoting buckling instabilities, meaning that this three-dimensional transition occurs for much lower values of $\elastohydronum$.

We note that, in these simulations, the hydrodynamics are forced by effectively applying a distribution of tangential slip flows (generated by the chemical activity) that are independent of time and fixed in the body frame. Since we are using a linear local slender body theory for Stokes flow --- resistive force theory --- identical dynamics may be obtained by instead applying a distribution of tangential forces along the centreline, since tangential flow and force are linearly related. However, more generally, we do not expect this equivalence to hold for more accurate models, for example using non-local theories for chemical activity or flow.

\begin{figure}
    \centering
\includegraphics[width=\linewidth]{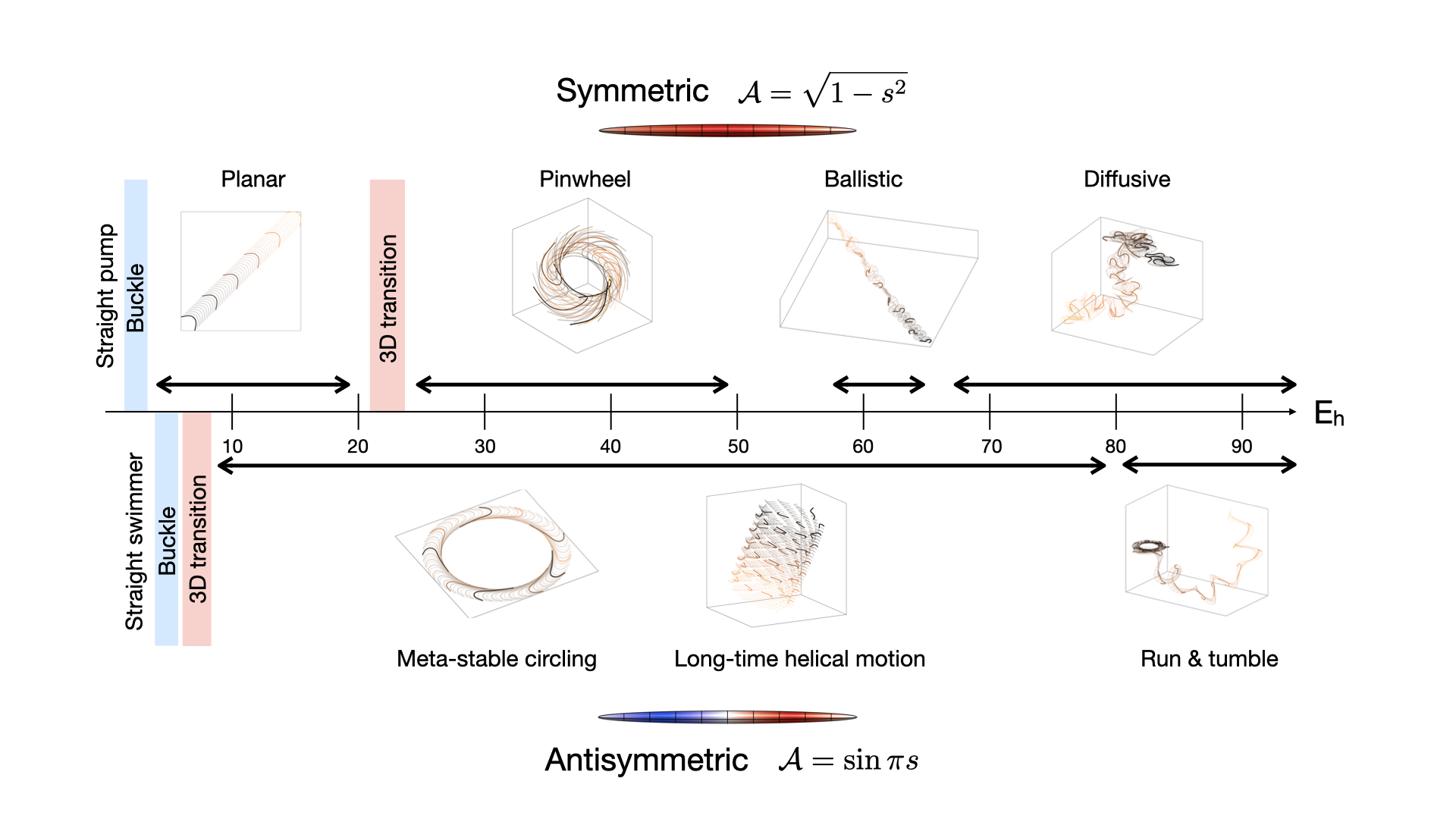}
    \caption{Categorisation of the observed dynamic behaviours of the symmetric and antisymmetric chemical activities. In each case, as $\elastohydronum$ increases, we observe a planar buckling transition, followed by an out-of-plane transition. Beyond this we see an array of dynamics, including closed orbits, helical trajectories, near-planar states and chaotic diffusion-like motion.}
    \label{fig:summaryfig}
\end{figure}

\section{Discussion} \label{sec:Discussion}

In this work, we have investigated the dynamics that arise when coupling the hydrodynamic loads induced by diffusiophoretic motility with the elastic behaviour of flexible filaments. 
In order to explore a large portion of the parameter space, we have opted for a coarse, rapid simulation methodology. This incorporates local versions of the slender body theories for hydrodynamic forcing and diffusion-dominated reactions, coupling resistive force theory with local slender phoretic theory. From our validations of local slender phoretic theory and previous validation of the approach to slender elastohydrodynamics, we expect that this provides good qualitative agreement and fair quantitative agreement with higher resolution methods. 
The key dimensionless quantity governing the strength of phoresis-induced hydrodynamic forces to elastic restoring forces, is the elastohydrodynamic number, $\elastohydronum$. By varying this parameter across orders of magnitude, we have uncovered a profoundly rich array of dynamic behaviours that arise from the interplay between phoresis, hydrodynamics and three-dimensional elasticity.

In this exploratory study, we focused on slender spheroidal filaments that naturally relax to straight configurations, and on two archetypical activity patterns: symmetric Saturn- and antisymmetric Janus-like catalytic coatings. 
In the rigid limit, these correspond to stationary pumps and self-propelling translators, respectively.  When elasticity is introduced, these baseline behaviours evolve into a wide spectrum of non-linear dynamical states. In both examples, as the elastohydrodynamic number is increased, equivalent to the filaments becoming more active or more flexible, the dynamics of the filaments pass through several key dynamic regimes. These regimes are summarised in \Cref{fig:summaryfig}.

At very low $\elastohydronum$, elasticity dominates activity and filaments relax towards their rigid behaviour. Once $\elastohydronum$ reaches a threshold value, we observe buckling phenomena that induce subsequent planar motion. This buckling is reminiscent of classical Euler buckling in beams and rods as the compressive load (here, due to viscous forcing) is increased. For Janus filaments, this buckling precipitates the onset of swimming behaviour; for the Saturn filaments, this transitions the dynamics from translation to rotation.
Increasing the elastohydrodynamic number further, early-time planar motion is seen to become unstable, and fully three-dimensional helical motion emerges. For the antisymmetric Saturn filaments, this occurred very soon after the buckling transition, and further study is needed to determine whether any planar states are stable on long-times for these filaments. We observe that the out-of-plane transition is typically preceded by a localised increase in torsion near the least curved section of the filament, and this instability may allow for the relaxation of elastic energies in highly curved filaments by deforming out-of plane.
At very high $\elastohydronum$, the filaments are easily deformed by the activity-induced viscous forces, leading to rapid shape changes that appear as diffusive-like motion. 

These types of behaviours are similar to those seen in other models of active and forced elastic filaments, which often show regimes of linear, planar and three-dimensional motion \citep[e.g.~][]{duRoure2019dynamics,laskar2013hydrodynamic,clarke2024bifurcations,altunkeyik2025dynamics}. However,
we note that the details of the motion depend sensitively on the nature of the activity/forcing and the applied boundary conditions. For instance, as the strength of activity is increased, clamped filaments with follower forces have been seen to exhibit an initial three-dimensional whirling motion before planar deformation \citep{clarke2024bifurcations}; instead we observe an initial planar instability, with three-dimensional motion occurring for higher values of activity. We expect that, in other systems, we shall see different ordering of modes, at different parameter values, with different shape transitions. Further study is needed to determine whether the behaviour seen here is characteristic of free-swimming elastic active filaments, or specific to chemoelastohydrodynamic filaments.

Although the general trends seen in both examples of activity profile were similar, including transitions from rigid to planar to helical and diffusive behaviours, there are some additional noteworthy dynamical regimes specific to each example at intermediate values of the elastohydrodynamic number. For example, there are a range of values of $\elastohydronum$, not far beyond the transition to three-dimensional motion, where we observed symmetric filaments settling into trapped periodic `pinwheeling' orbits. These orbits emerge during the course of the motion, and appear to be stable attractors. We also observe a finite-width region of parameter space where the overall trajectories return to near-planar ballistic motion, with travelling waves moving from the filament centre to the tip inducing a flapping-like motion from side-to-side. Further study is required to understand more fully why these dynamics emerge, along with the short or long-time persistence of apparent transients.

This study is a key step in understanding the behaviour of free-swimming chemically propelled elastic filaments.
The results presented here are numerical experiments of an initial-value problem, and not a full analysis of the nonlinear dynamics. Hence, this study can be viewed as experimental evidence of the dynamics of these theoretical systems. Although we have made attempts to check the robustness of the presented behaviours to different conditions, it is possible that new dynamics may emerge, particularly on longer timescales. As such, we advocate for a detailed stability analysis of the dynamics, though this may be complicated by the non-small deviation of the dynamics from the elastic equilibrium. Such an analysis would likely be especially fruitful in the context of the antisymmetric activity profile considered in this work, where emergent behaviours were observed to depend on initial conditions.

Our numerics rely on asymptotically derived theories for both the chemical concentration and viscous flow around a thin and long filament. In each case, we use a local slender body theory, which leads to typical errors of a moderate size, $O(1/\log\epsslend)$, as discussed in \cref{app:local_vs_nonlocal}. The errors are likely to be particularly acute when initially distant parts of the filaments become close together and non-local effects are more significant. In our simulations, this appears to occur when the filament is at its most deformable, such as the examples shown in \cref{fig:beyond_helix}; this suggests that there may be higher errors associated with the results for extremely high values of $\elastohydronum$. We otherwise expect that our simulations provide reasonable accuracy for the dynamic behaviours observed across the wide range of parameter values considered.

We envision this initial exploration of chemoelastohydrodynamic filaments as the doorstep along several exciting directions. Methodological advances in numerical simulations of chemically active filaments will allow for high fidelity and high accuracy exploration of the dynamic behaviours described. Our simulations have revealed an intricate array of dynamic transitions, and detailed explorations of these are already commencing, including using linear stability analysis and spectral methods to investigate the initial onset of buckling and subsequent swimming of uniformly active phoretic filaments \citep{makanga2025instability}.
Further exploration of the parameter space is sure to reveal exciting behaviours when additional degrees of freedom are activated, such as non-axisymmetric activity patterns or other spatial, temporal or stochastic variations in surface activity.  
Moreover, the filament dynamics in more complex fluid environments, such as viscoelastic or shear flows, and their interactions in suspensions offer rich grounds for cross-fertilisation with active rheology and collective behaviour studies by the active soft matter community. 
Coupling the present framework with stimuli-responsive materials and control theory will open applications for controllable microbot navigation and microfluidic mixing \citep{MicrotransformersTom2018}.
We also hope this work stimulates further experimental advances in the fabrication of flexible phoretic filaments. 

\section*{Acknowledgements} 
MDB was funded by a Clifford Fellowship at University College London. BJW is supported by the Royal Commission for the Exhibition of 1851.  TDMJ and PK gratefully acknowledge funding from the Engineering
and Physical Sciences Research Council (EPSRC) Grant No. EP/R041555/1 “Artificial
Transforming Swimmers for Precision Microfluidics Tasks” to TDMJ.  TDMJ gratefully acknowledges funding from the Leverhulme Trust Research Leadership
Award “Shape-Transforming Active Microfluidics” Grant No. RL-2019-014. PK also acknowledges the support from the Cyprus Research and Innovation Foundation under contract No.~EXCELLENCE/0524/0363 for the Excellence Hub project `MICROFIBRES' to PK and expresses gratitude to Dr Christos Marangos for support in figure 1 schematic.  
The authors are grateful to Dr Ursy Makanga and Dr Akhil Varma for discussions on the subject. 

\section*{Declaration of Interests} 
The authors report no conflict of interest.

\begin{appendix}

\section{Non-local SPT} \label{app:nonlocalSPT}

For an axisymmetric activity, $\activity(s)$, the first two terms of the surface concentration expansion for \emph{non-local} slender phoretic theory, were previously calculated as \citep{SPT2020}
\begin{align}
        \zerothorder{c}(s)
        &= 
        \frac{1}{2}
        \crossradius(s)\activity(s)\log\left(\frac{4 }{\epsslend^2 \crossradius^2(s)}\right)
        +
        \frac{1}{2} \int_{-1}^{1} \left[
        \frac{\crossradius(\sdum)\activity(\sdum)}{ |\bvRo(s,\sdum)|} 
        -\frac{\crossradius(s)\activity(s)}{|s-\sdum|} \right]
         \dd \sdum 
        \label{eq:ZerothOrderConc_AxiActivity}
        \\
        \firstorder{c}(s,\theta)
        &= 
             \frac{1}{2} \crossradius^2(s) \curvature(s) \activity(s)	
            \cos\Theta(s,\theta)  \left[\log\left(\frac{4}{\epsslend^2 \crossradius^2(s)}  \right)
            - 3\right]
        \nonumber
        \\
        &\qquad - \crossradius(s) \int\limits_{-1}^{1}\!
        \left[ 
        \frac{\crossradius(\sdum)\activity(\sdum)}{ |\bvRo(s,\sdum)|^3}    
              \bvRo(s,\sdum)    \cdot\erho(s,\theta)
          +  \frac{\crossradius (s)\curvature(s)\activity(s)\cos\Theta(s,\theta)}{2|s-\sdum|}\right]\dd \sdum  , 
          \label{eq:FirstOrderConc_AxiActivity}
\end{align}
where $\erho(s,\theta)$ is the unit vector pointing radially outwards from any cross-section. 
The \emph{local} version of slender phoretic theory, used in this work, neglects the line integral contributions as they are size $\log(1/\epsslend)$ smaller than the logarithmic terms.

\section{Verifying local SPT} \label{app:local_vs_nonlocal}

\subsection{Local vs non-local SPT}

\begin{figure}
    \centering
    \begin{tikzpicture}
        \node[anchor = west,font = \large] at (0,0) {(a)};
        \node[anchor = west,font = \large] at (7.5,0) {(b)};
        \node[anchor = west] at (15,0) {};
    \end{tikzpicture}\\
    \includegraphics[width=0.9\linewidth]{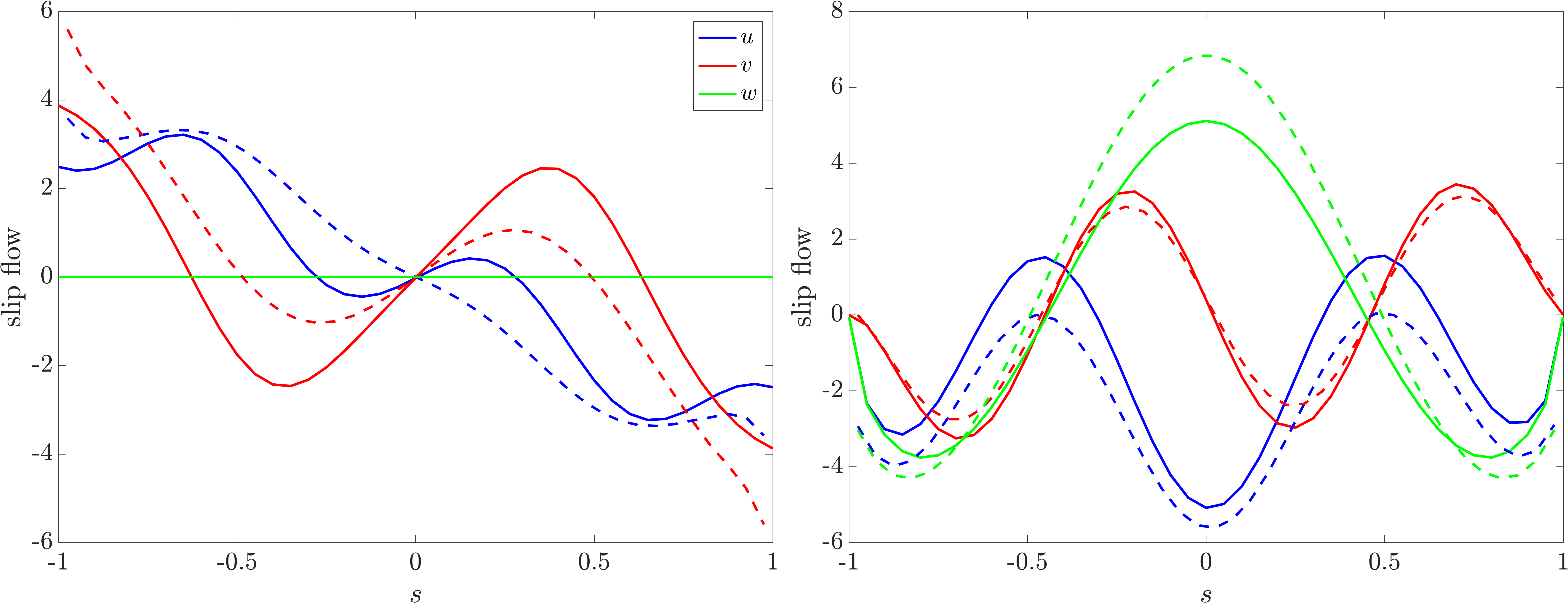}
    \caption{Comparison of leading order slip flows calculated by local (solid) and non-local (dashed) SPT. (a) An S-shaped filament with $\activity=\sqrt{1-s^2}$, (b) a helical filament with $\activity=\sin{\pi s}$. In both cases, we used a mobility $\mobility=+1$, with $N=40$ segments and slenderness $\epsslend=0.02$. }
    \label{fig:local_vs_nonlocal}
\end{figure}

We verify the accuracy of using a local version of slender phoretic theory by comparing the calculated slip flows with those computed by the non-local theory, which has been previously validated \citep{SPT2020,katsamba2024slender}. An example of the calculated slip flows for each of our considered activity profiles are shown in \Cref{fig:local_vs_nonlocal}, for a fixed prescribed shape, showing good agreement between the local and non-local theories at the slenderness considered. 

\subsection{Convergence study}

\begin{figure}
    \centering
    \begin{tikzpicture}
        \node[anchor = west,font = \large] at (0,0) {(a)};
        \node[anchor = west,font = \large] at (7.3,0) {(b)};
        \node[anchor = west] at (15,0) {};
    \end{tikzpicture} \\
    \includegraphics[width=0.9\linewidth]{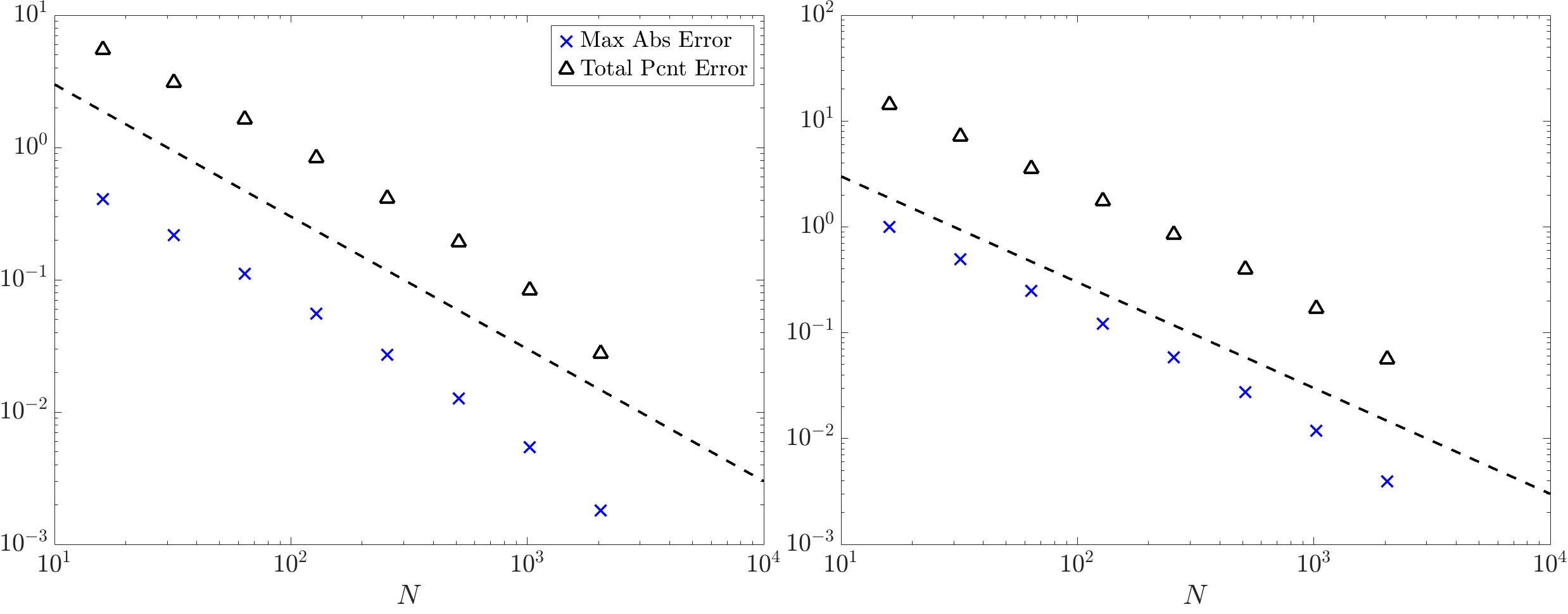}
    \caption{Convergence of the local SPT code. The instantaneous slip flow from local SPT is calculated for different number of segments, $N=2^n$, in the discretisation, and compared to the highest resolution, $m=12$.
    The given shape and activity are:  (a) an S-shape with $\activity=\sqrt{1-s^2}$, and (b) a helical filament with $\activity = \sin(\pi s)$. 
    Blue crosses denote the maximum absolute error in slip velocity, black triangles are the percentage error over the entire filament, calculated as $\epsilon(N)=(\sum_{segs} |U_N-U_{\mathrm{ref}}|)/(\sum_{segs} |U_{\mathrm{ref}}|)$ expressed as a percentage.
    The dashed line decreases like $N^{-1}$. 
    }
    \label{fig:SPTconvergence}
\end{figure}

To further confirm the reliability and accuracy of our local slender phoretic theory, we calculated the instantaneous slip flows generated by filaments, each with a given shape and activity, as the number of discretised segments was varied. Two examples of this are given in \Cref{fig:SPTconvergence}, showing the error decreases inversely proportional to the number of segments. Our choice of $N=40$ for our simulations, suggests that we can calculate the slip velocity to within 10\% accuracy, which is consistent with the error of resistive force theory used for the resulting hydrodynamics.

\end{appendix}

\newpage

\bibliographystyle{jfm}
\bibliography{main}

\end{document}

%% file: SPTmacro.tex
\newcommand{\bv}[1]{\bm{#1}} 
\renewcommand{\vec}[1]{\bv{#1}}
\newcommand{\dd}{\mathrm{d}} 

\newcommand{\activity}{\mathcal{A}}
\newcommand{\mobility}{\mathcal{M}}
\newcommand{\vslip}{\bv{v}_{\mathrm{slip}}}
\newcommand{\no}{\bv{n}_f}
\newcommand{\idmat}{\bv{1}}
\newcommand{\x}{\bv{x}}
\newcommand{\activityscale}{A}
\newcommand{\mobilityscale}{M}

\newcommand{\zerothorder}[1]{{#1}^{(0)}}
\newcommand{\firstorder}[1]{{#1}^{(1)}}

\newcommand{\sdum}{\tilde{s}}
\newcommand{\thetadum}{\tilde{\theta}}
\newcommand{\length}{l}

\newcommand{\erho}{\hat{\bv{e}}_{\rho}}
\newcommand{\etheta}{\hat{\bv{e}}_{\theta}}
\newcommand{\tanhat}{\hat{\bv{t}}}
\newcommand{\norhat}{\hat{\bv{n}}}

\newcommand{\crossradius}{\rho}
\newcommand{\maxcrossradius}{r_f}

\newcommand{\curvature}{\kappa}

\newcommand{\thetamode}{\Theta(s,\theta)}

\newcommand{\bvRo}{\bv{R}_0}
\newcommand{\epsslend}{\epsilon}  

\newcommand{\coeffsinmode}{A_s}

\newcommand{\darboux}{\bv{d}} 
\newcommand{\cdarboux}[1]{\darboux_{#1}}

\newcommand{\Finternal}{\mathbf{F}}
\newcommand{\Mbending}{\mathbf{M}}
\newcommand{\forcedens}{\bv{f}}
\newcommand{\torquedens}{\bv{\tau}}

\newcommand{\uparallel}{\bv{u}_\parallel}
\newcommand{\uperp}{\bv{u}_\perp}

\newcommand{\Xc}{\bv{X}}

\newcommand{\vslipo}{\bar{\bv{v}}_{slip}}

\newcommand{\elastohydronum}{E_h}